\documentclass[journal]{IEEEtran}

\usepackage{graphicx,xcolor}
\DeclareGraphicsExtensions{.eps,.jpeg,.png}
\usepackage{amsmath}
\usepackage{amsfonts}

\usepackage{array}
\usepackage{amssymb,mathtools}
\usepackage{empheq}
\usepackage{epstopdf}
\usepackage[linesnumbered,ruled]{algorithm2e}

\usepackage{array}
\usepackage{caption}
\usepackage{subcaption}

\usepackage{bm,amsmath,amssymb,amsfonts,graphicx,epsfig,amsthm,color}
\usepackage[hyphens]{url}
\PassOptionsToPackage{bookmarks=false}{hyperref}

\usepackage[depth=-1]{bookmark}

\usepackage{times}
\usepackage{graphicx} 
\usepackage{subcaption} 

\usepackage{wrapfig}

\usepackage{accents}
\makeatletter
\def\wid{\check{{\cc@style\underline{\mskip9.5mu}}}}
\def\Wideubar{\underaccent{{\cc@style\underline{\mskip6mu}}}}
\makeatother

\makeatletter
\def\wideubar{\underaccent{{\cc@style\underline{\mskip9.5mu}}}}
\def\Wideubar{\underaccent{{\cc@style\underline{\mskip6mu}}}}
\makeatother

\makeatletter
\def\widebar{\accentset{{\cc@style\underline{\mskip9.5mu}}}}
\def\Widebar{\accentset{{\cc@style\underline{\mskip6mu}}}}
\makeatother

\theoremstyle{remark}\newtheorem{remark}{Remark}

\allowdisplaybreaks

\begin{document}
	
	\title{Deep Reinforcement Learning for Adaptive Caching  
		in Hierarchical Content Delivery Networks}
	\author{Alireza Sadeghi, Gang Wang, and Georgios B. Giannakis
		\thanks{This work was supported in part by NSF grants 1711471 and 1514056. The authors are with the Digital Technology Center and the Department of Electrical and Computer Engineering, University of Minnesota, Minneapolis, MN 55455, USA. E-mails: \{sadeghi, gangwang, georgios\}@umn.edu. 
		}
	}
	
	\maketitle
	\begin{abstract}
		Caching is envisioned to play a critical role in next-generation content delivery infrastructure, cellular networks, and Internet architectures. By smartly storing the most popular contents at the storage-enabled network entities during off-peak demand instances, caching can benefit both network infrastructure as well as end users, during on-peak periods. In this context, distributing the limited storage capacity across network entities calls for decentralized caching schemes. Many practical caching systems involve a parent caching node connected to multiple leaf nodes to serve user file requests. To model the two-way interactive influence between caching decisions at the parent and leaf nodes, a reinforcement learning (RL) framework is put forth. To handle the large continuous  state space, a scalable deep RL approach is pursued. The novel approach relies on a hyper-deep Q-network to learn the Q-function, and thus the optimal caching policy, in an online fashion. Reinforcing the parent node with ability to learn-and-adapt to unknown policies of leaf nodes as well as spatio-temporal dynamic evolution of file requests, results in remarkable caching performance, as corroborated through numerical tests.     
	\end{abstract}
	
	\section{Introduction}
	\label{Sec:Intro}
	In light of the tremendous growth of data traffic over both wireline and wireless communications, next-generation networks including future Internet architectures, content delivery infrastructure, and cellular networks stand in need of emerging technologies to meet the ever-increasing data demand. Recognized as an appealing solution is \textit{caching}, which amounts to storing reusable contents in geographically distributed storage-enabled network entities so that future requests for those contents can be served faster. The rationale is that unfavorable shocks of peak traffic periods can be smoothed by proactively storing `anticipated' highly popular contents at those storage devices and during off-peak periods~\cite{Paschos18, Bastug14}. Caching popular content is envisioned to achieve major savings in terms of energy, bandwidth, and cost, in addition to user satisfaction~\cite{Paschos18}.
	
To fully unleash its potential, a content-agnostic caching entity has to rely on available observations to learn what and {when} to cache. Toward this goal, contemporary machine learning and artificial intelligence tools hold the promise to empower next-generation networks with `smart' caching control units, that can learn, track, and adapt to unknown dynamic environments, including space-time evolution of content popularities and network topology, as well as entity-specific caching policies.

Deep neural networks (DNNs) have lately boosted the notion of ``learning from data'' with field-changing performance improvements reported in diverse  artificial intelligence tasks \cite{goodfellow2016deep}. DNNs can cope with the `curse of dimensionality' by providing compact low-dimensional representations of high-dimensional data~\cite{pami2013bengio}. Combining deep learning with RL, deep (D) RL has created the first artificial agents to achieve human-level performance across many challenging domains~\cite{minh2015,survey2019}. As another example, a DNN system was built to operate Google's data centers, and shown able to consistently achieve a 40\% reduction in  energy consumption for cooling \cite{2014datacenter}. This system provides a general-purpose framework to understand complex dynamics, which has also been applied to address other challenges {\color{black} including e.g., dynamic spectrum access \cite{DRLSA}, multiple access and handover control \cite{DRLMA}, \cite{DRLHO}, as well as resource allocation in fog-radio access networks \cite{DRLMS}. }

\subsection{Prior art on caching}
Early approaches to caching include the least recently used (LRU), least frequently used (LFU), first in first out (FIFO), random replacement (RR) policies, and their variants. Albeit simple, these schemes cannot deal with the dynamics of content popularities and network topologies.	Recent efforts have gradually shifted toward developing learning and optimization based approaches that can `intelligently' manage the cache resources. For unknown but time-invariant content popularities, multi-armed bandit online learning (e.g., \cite{2018lbc}) was pursued in~\cite{multiarm2014}. {\color{black} Yet, these methods are generally not amenable to online implementation. To serve delay-sensitive requests, a learning approach was developed in~\cite{DNN_NonCVX} using a pre-trained DNN to handle a non-convex problem reformulation.}
	
In realistic networks however, popularities exhibit dynamics, which motivate well the so-termed \emph{dynamic} caching. A Poisson shot noise model was adopted to approximate the evolution of popularities in \cite{PSN1}, for which an age-based caching solution was developed in~\cite{PSN2}. RL based methods have been pursued in \cite{RL1, RL2, RL3, CacheIA}. Specifically, a {Q}-learning based caching scheme was developed in \cite{RL1} to model global and local content popularities as Markovian processes. Considering Poisson shot noise popularity dynamics, a policy gradient RL based caching scheme was devised in \cite{RL2}. Assuming stationary file popularities and service costs, a dual-decomposition based Q-learning approach was pursued in~\cite{RL3}. {\color{black} Albeit reasonable for discrete states, these approaches cannot deal with large continuous state-action spaces.} {\color{black}To cope with such spaces, DRL approaches have been considered for content caching in e.g., \cite{CacheIA, DRL_AC, DRL_Vehic, DRL_Smartcity,survey2019, DRL4edgecaching}. Encompassing finite-state time-varying Markov channels, a deep Q-network approach was devised in~\cite{CacheIA}. An actor-critic method with deep deterministic policy gradient updates was used in~\cite{DRL_AC}. Boosted network performance using DRL was documented in several other applications, such as connected vehicular networks~\cite{DRL_Vehic}, and smart cities \cite{DRL_Smartcity}.}

The aforementioned works focus on devising caching policies for a \emph{single} caching entity. A more common setting in next-generation networks however, involves a network of interconnected caching nodes. It has been shown that considering a network of connected caches jointly can further improve performance \cite{Maddahali2014, Distributed2010}. For instance, leveraging network topology and the broadcast nature of links, the coded caching strategy in~\cite{Maddahali2014} further reduces data traffic over a network. This idea has been extended in \cite{Online2015} to an online setting, where popularities are modeled Markov processes. Collaborative and distributed online learning approaches have been pursued \cite{collaborative2012,Distributed2010,decentralized2018}. {\color{black} Indeed, today's content delivery networks such as Akamai \cite{nygren2010akamai}, have tree network structures. Accounting for the hierarchy of caches has become a common practice in recent works; see also \cite{ramadan2019framework, JointDehgan17,JointShukla18, tong2016hierarchical}. Specifically, a general framework to evaluate caching policies in a hierarchical network of caches was developed in \cite{ramadan2019framework}. Joint routing and in-network content caching in a hierarchical cache network was formulated in \cite{JointDehgan17}, for which greedy schemes with provable performance guarantees can be found in~\cite{JointShukla18}. }
	
We identify the following challenges that need to be addressed when designing practical caching methods for next-generation networks.
\begin{enumerate}
	\item[\bf{c1)}] \emph{Networked caching.} Caching decisions of a node, in a network of caches, influences decisions of all other nodes. Thus, a desired caching policy must adapt to the network topology and policies of neighboring nodes.

	\item[\bf{c2)}] \emph{Complex dynamics.} Content popularities are random and exhibit unknown space-time, heterogeneous, and often non-stationary dynamics over the entire network. 
	
	\item[\bf{c3)}] \emph{Large continuous state space.} Due to the shear size of available content, caching nodes, and possible realizations of content requests, the decision space is huge. 
		
	\end{enumerate} 
	
\subsection{This work} 
{\color{black} Prompted by the recent interest in hierarchical caching, this paper focuses on a two-level network caching, where a parent node is connected to multiple leaf nodes to serve end-user file requests. Such a two-level network constitutes the building block of the popular tree hierarchical cache networks in e.g., \cite{ramadan2019framework, nygren2010akamai}. To model the interaction between caching decisions of parent and leaf nodes along with the space-time evolution of file requests, a scalable DRL approach  based on hyper deep Q-networks (DQNs) is developed. As corroborated by extensive numerical tests, the novel caching policy for the parent node can adapt itself to local policies of leaf nodes and space-time evolution of file requests. Moreover, our approach is simple-to-implement, and performs close to the optimal policy. }

\begin{figure}
		\centering
		\includegraphics[width =0.2 \textwidth]{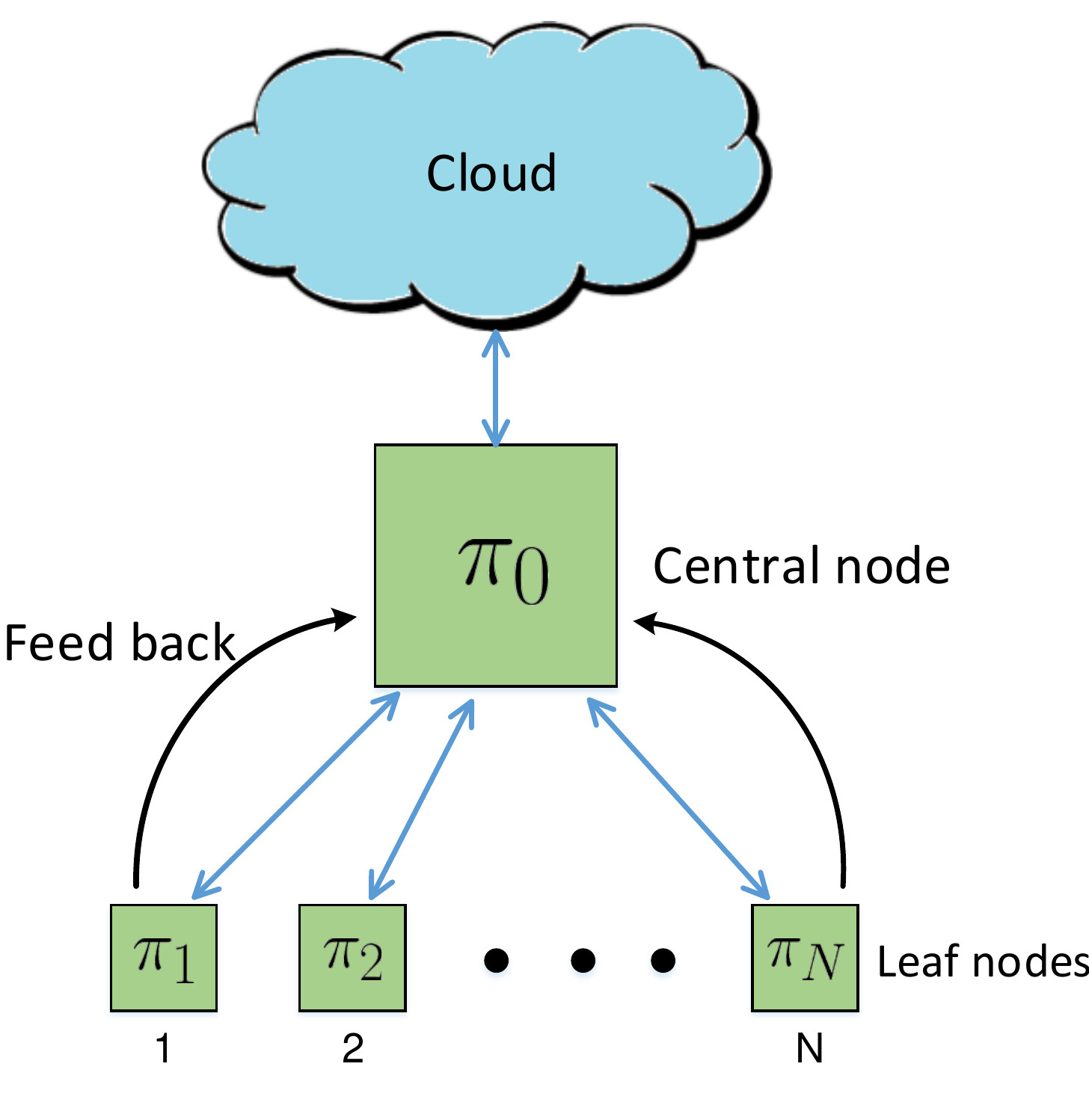}
		\caption{A network of caching nodes.}
		\label{fig:model}
		\centering \includegraphics[width =0.22\textwidth]{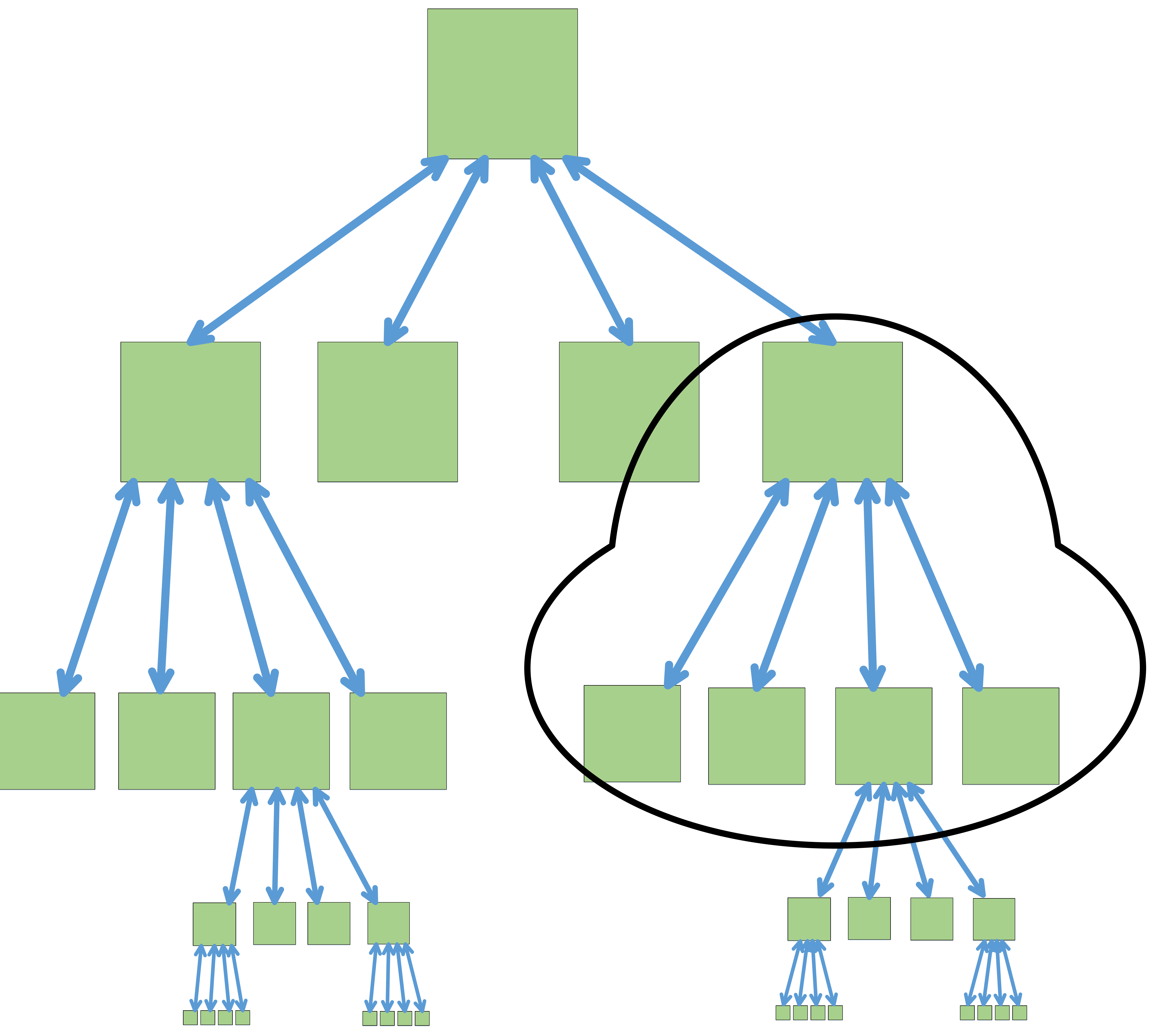}	\caption{A hierarchical tree network cache system.} 
		\label{fig:higherarchy}
\end{figure}
	
\section{Modeling and Problem Statement}\label{sec:model}
	
Consider a two-level network of interconnected caching nodes, where a parent node is connected to $N$ leaf nodes, indexed by $n\in\mathcal{N}:=\{1,\ldots,N\}$. The parent node is connected to the cloud through a (typically congested) back-haul link; see Fig.~\ref{fig:model}. One could consider this network as a part of a large hierarchical caching system, where the parent node is connected to a higher level caching node instead of the cloud; see Fig.~\ref{fig:higherarchy}. In a content delivery network for instance, edge servers (a.k.a. points of presence or PoPs) are the leaf nodes, and a fog server acts as the parent node. Likewise, (small) base stations in a 5G cellular network are the leaf nodes, while a serving gate way (S-GW) may be considered as the parent node; see also~\cite[p.~110]{LTE}.   
	
All nodes in this network store files to serve file requests. Every leaf node serves its locally connected end users, by providing their requested files. If a requested content is locally available at a leaf node, the content will be served immediately at no cost. If it is not locally available due to limited caching capacity, the content will be fetched from its parent node, at a certain cost. Similarly, if the file is available at the parent node, it will be served to the leaf at no cost; otherwise, the file must be fetched from the cloud at a higher cost.
	
To mitigate the burden with local requests on the network, each leaf node stores `anticipated' locally popular files. In addition, this paper considers that each parent node stores files to serve requests that are \emph{not} locally served by leaf nodes. Since leaf nodes are closer to end users, they frequently receive file requests that exhibit rapid temporal evolution at a \emph{fast} timescale. The parent node on the other hand, observes aggregate requests over a large number of users served by the $N$ leaf nodes, which naturally exhibit smaller fluctuations and thus evolve at a \emph{slow} timescale. 

\begin{figure}
		\centering
		\includegraphics[width =0.46 \textwidth]{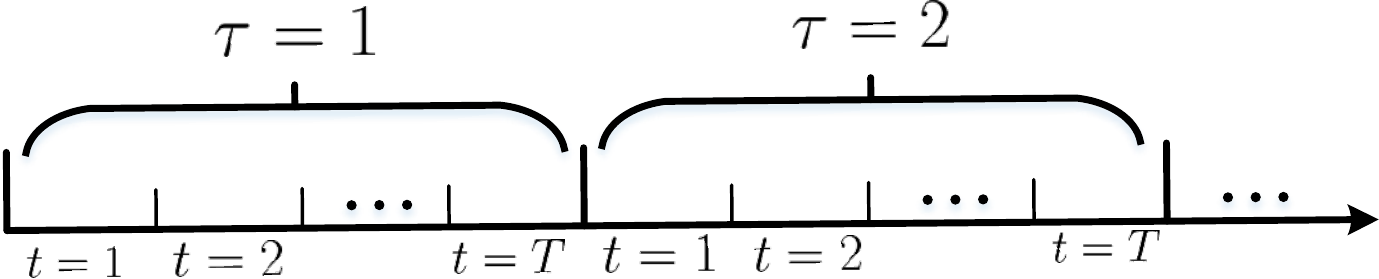}
		\caption{Slow and fast time slots.}
		\label{fig:Timescales}
						\vspace{-5pt}
\end{figure}

This motivated us to pursue a two-timescale approach to managing such a network of caching nodes. To that end, let $\tau=1,2,\ldots$ denote the slow time intervals, each of which is further divided into $T$ fast time slots indexed by $t=1,\ldots,T$; see Fig.~\ref{fig:Timescales} for an illustration. Each fast time slot may be e.g., $1$-$2$ minutes depending on the dynamics of local requests, while each slow time interval is a period of say $4$-$5$ minutes. We assume that the network state remains unchanged during each fast time slot $t$, but can change from $t$ to $t+1$.
	
Consider a total of $F$ files in the cloud, which are collected in the set ${\mathcal F} = \{1,  \ldots, F\}$. At the beginning of each slot $t$, 	every leaf node $n$ selects a subset of files in $\mathcal{F}$ to prefetch and store for possible use in this slot.
To determine which files to store, every leaf node relies on a local caching policy function denoted by $\pi_n$, to take (cache or no-cache) action ${\pmb a}_n ( t+1, \tau  ) = \pi_n (\pmb s_n (t, \tau)) $ at the beginning of slot $t+1$, based on its \emph{state} vector ${\pmb s}_n$ at the end of slot $t$. We assume this action takes a negligible amount of time relative to the slot duration; and  define the state vector $\pmb s_n (t, \tau)\! :=\! \pmb r_n(t,\tau) := [r_n^1(t,\tau) \cdots r_n^{F}(t,\tau)]^\top$ to collect the number of requests received at leaf node $n$ for individual files over the duration of slot $t$ on interval $\tau$. Likewise, to serve file requests that have not been served by leaf nodes, the parent node takes action $\pmb a_0 ( \tau )$ to store files at the beginning of every interval $\tau$, according to a certain policy $\pi_0$. Again, as aggregation smooths out request fluctuations, the parent node observes slowly varying file requests, and can thus make caching decisions at a relatively slow timescale. In the next section, we present a two-timescale approach to managing such a network of caching nodes.

\section{Two-timescale Problem Formulation}\label{sec:two}
File transmission over any network link consumes resources, including e.g., energy, time, and bandwidth. Hence, serving any requested file that is not locally stored at a node, incurs a cost. Among possible choices, the present paper considers the following cost for node $n\in\mathcal{N}$, at slot $t+1$ of interval $\tau$
\begin{multline}
\hspace{-0.3 cm} \pmb c_{n}(\pi_{n} (\pmb s_n ( t,\tau ) ), \pmb r_{n} (t+1,\tau), \pmb a_0 (	\tau ) )  \! :=  \! \pmb r_n (t+1, \tau) \odot  ( {\bf 1}\! - \pmb a_{0} (\tau))  \\  \!\odot \! ( {\bf 1} \!- \pmb a_{n} (t\!+\!1,\tau) )\! + \pmb r_n (t\!+\!1, \tau) \odot ( {\bf 1} \!- \pmb a_n(t\!+\!1,\tau) )
\label{eq:nodalcost}
\end{multline}
where $\pmb c_{n} (\cdot):=[c_n^1(\cdot)~\cdots~c_n^F(\cdot)]^{\top}$ concatenates the cost for serving individual files per node $n$; symbol $\odot$ denotes entry-wise vector multiplication; entries of $\pmb a_0$ and $\pmb a_n$ are either $1$ (cache, hence no need to fetch), or, $0$ (no-cache, hence fetch); and $\bf 1$ stands for the all-one vector. Specifically, the second summand in \eqref{eq:nodalcost} captures the cost of the {\color{black} leaf} node fetching files for end users, while the first summand corresponds to that of the parent fetching files from the cloud. 

We model user file requests as Markov processes with unknown transition probabilities~\cite{RL1}. Per interval $\tau$, a reasonable caching scheme for leaf node $n\in\mathcal{N}$ could entail minimizing the expected cumulative cost; that is, 
	\begin{align} 
	\pi^{ \ast}_{n,\tau} \!\!:= \underset{\pi_n \in \Pi_n}{\arg \min} ~ {\mathbb{E}}\Big[ \!\sum_{t=1}^{T}\! {\bf 1}^{\!\top} \!\pmb c_{n} (\pi_{n} (\pmb s_n ( t,\tau ) ), \pmb r_{n} (t\!+\!1,\tau), \pmb a_0 (\tau ))\Big] 
	\label{eq:leaf_node}
	\end{align} 
where $\Pi_n$ represents the set of all feasible policies for node~$n$. Although solving \eqref{eq:leaf_node} is in general challenging, efficient near-optimal solutions have been introduced in several recent contributions; see e.g.,~\cite{RL1,RL2,multiarm2014}, and references therein. In particular, a RL based approach using tabular $Q$-learning was pursued in our precursor \cite{RL1}, which can be employed here to tackle this fast timescale optimization. The remainder of this paper will thus be on designing the caching policy $\pi_0$ for the parent node, that can learn, track, and adapt to the leaf node policies as well as user file requests.   
	
\begin{figure}
		\centering
		\includegraphics[width =0.3 \textwidth]{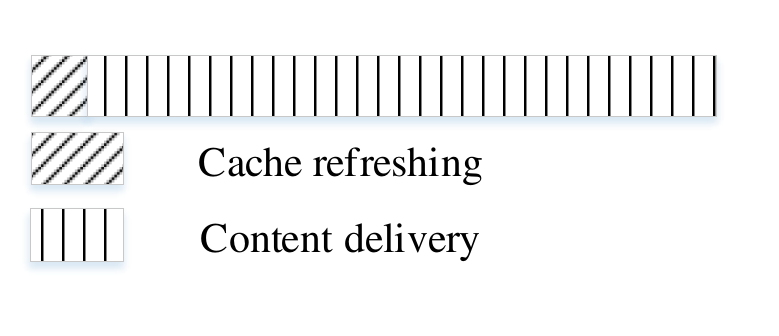}
				\vspace{-5pt}
		\caption{Structure of slots and intervals.}
		\label{fig:slotstructure}
\end{figure}
	
At the beginning of every interval $\tau +1$, the parent node takes action to refresh its cache; see Fig.~\ref{fig:slotstructure}. To design a caching policy that can account for both leaf node policies and file popularity dynamics, the parent collects local information from all leaf nodes. At the end of interval $\tau$, each leaf node $n\in\mathcal{N}$ first obtains the time-average state vector ${\bar {\pmb s}_n} (\tau):= (1/T) \sum_{t=1}^{T}  \pmb s_{n} ( t , \tau)$, and subsequently forms and forwards the per-node vector ${\bar {\pmb s}_{n}} ( \tau ) \odot ({\bf 1} - \pi_n ( {\bar {\pmb s}_{n}} ( \tau )) )$ to its parent node. This vector has nonzero entries the average number of file requests received during interval $\tau$, and zero entries if $\pi_n$ stores the corresponding files. Using the latter, the parent node forms its `weighted' state vector as
	\begin{equation}
	\label{eq:netstate}
	\pmb s_0 (\tau):= \sum_{n = 1}^{N} w_n {\bar {\pmb s}_{n}} ( \tau ) \odot ({\bf 1} - \pi_n \!\left( {\bar {\pmb s}_{n}} (\tau) \right) )
	\end{equation}
where the weights $\{w_n\ge 0\}$ control the influence of leaf nodes $n\in\mathcal{N}$ on the parent node's policy. Similarly, having received at the end of interval $\tau +1$ the slot-averaged costs
\begin{multline}
	{\bar {\pmb c}_n}( \pi_0 (\pmb s_0 ( \tau))  ;  \{ \pmb a_n ( t, \tau +1), \pmb r_n ( t , \tau +1) \}_{t=1}^T )\\ 
	 = \frac{1}{T} \sum_{t = 1}^{T} \pmb c_{n}(\pmb a_{n} (t,\tau+1), \pmb r_{n} \!\left(t,\tau+1 \right), \pmb a_0 (	\tau +1) )\label{eq:cost_pernode}
\end{multline}
from all leaf nodes, the parent node determines its cost 
\begin{align} \label{eq:c0}
& \pmb c_0 ( {\pmb s}_0 ( \tau), \pi_0 (\pmb s_0 ( \tau)) )   \\ \nonumber &  = \sum_{n=1}^{N} w_n  {\bar {\pmb c}_n}( \pmb \pi_0 ( \pmb s_0 (\tau) )  ; \{ \pmb a_n ( t, \tau +1), \pmb r_n ( t , \tau +1) \}_{t=1}^T ) ). 
\end{align}
Having observed $\left\{\pmb s_0( \tau' ), \pmb a_0 ( \tau'), \pmb c_0 ( {\pmb s}_0 ( \tau'-1), \pmb a_0 (\tau') \right\}_{\tau' = 1}^{ \tau}$, the objective is to take a well-thought-out action $\pmb a_0 ( \tau + 1 )$ for next interval. This is tantamount to finding an optimal policy function $\pi^{\ast}_0$ to take a caching action $\pmb a_0 \left(\tau \!\! +\! 1 \right) = \pi^{\ast}_0 \left(\pmb s_0 \left( \tau \right)\right)$. 
	
Since the requests $\{\pmb r_n (t,\tau)\}_{n,t}$ are Markovian and present actions $\{\pmb a_n ( t , \tau )\}_{n,t}$ also affect future costs, the optimal RL policy for the parent node will minimize the expected cumulative cost over all leaf nodes in the long term, namely
\begin{equation}\label{eq:pi00}
	\pi_0^\ast:=\underset{\pi_0 \in \Pi_0} {\arg\min} \; \mathbb {E} \Big[ \sum_{\tau= 1}^{\infty} \gamma^{\tau-1} {\bf 1}\!^{\top} \! \pmb c_0\! \left( \pmb s_0 (\tau), \pi_0 \! \left( \pmb s_0 (\tau) \right)  \right)  \Big]
\end{equation}  
where $\Pi_0$ represents the set of all feasible policies; 
expectation is over $\{\pmb r_n (t,\tau)\}_{n,t}$, as well as (possibly random) parent $\pmb a_0 (\tau)$ and leaf actions $\{\pmb a_n ( t , \tau )\}_{n,t}$; and the discount factor $\gamma \in [ 0, 1)$ trades off emphasis of current versus future costs. 
	
It is evident from \eqref{eq:pi00} that the decision taken at a given state $\pmb s_0 (\tau)$, namely $\pmb a_0 (\tau +1) = \pi (\pmb s_0(\tau))$, influences the next state $\pmb s_0(\tau+1)$ through $\pi_{n,\tau} (\cdot)$  in \eqref{eq:netstate}, as well as the cost $\pmb c_0 (\cdot)$ in \eqref{eq:c0}. Therefore, problem \eqref{eq:pi00} is a discounted infinite time horizon Markov decision process (MDP). Finding the optimal policy of an MDP is NP-hard \cite{Tsitsiklis}. To cope with this complexity of solving 
\eqref{eq:pi00}, an adaptive  RL approach is pursued next. 

\section{Adaptive RL-Based Caching} \label{sec:DRL}
	
RL deals with action-taking policy function estimation in an environment with dynamically evolving states, so as to minimize a long-term cumulative cost. By interacting with the environment (through successive actions and observed states and costs), RL seeks a policy function (of states) to draw actions from, in order to minimize the average cumulative cost as in  \eqref{eq:pi00}~\cite{RLbook}. To proceed, we start by giving some basics on Markov decision processes (MDPs) \cite[p.~310]{RLbook}.

{\color{black}
\subsection{MDP formulation}
MDPs provide a well-appreciated model for decision making tasks. It is represented by a tuple $\left<{\mathcal S}, {\mathcal A}, {\mathcal C}, {\mathcal P}\right>$, where ${\mathcal S}$ denotes a set of possible states, $\mathcal A$ a set of feasible actions, $\mathcal C$ a set of costs, and $\mathcal P$ the set of state transition probabilities. In our problem of interest, at the end of each interval $\tau$, the parent node is in state ${\pmb s}_0(\tau) \in {\mathcal S}$ and takes a caching decision $\pmb a_0 (\tau+1) \in {\mathcal A}$ for the next interval $\tau+1$, according to its local caching policy $\pmb a_0 (\tau+1) = \pi_0{(\pmb s_0(\tau))}$. Upon proceeding to interval $\tau+1$, every leaf node $n$ serves its locally connected users. Let vector $\pmb r_n (t,\tau+1)$ collect the number of received requests at node $n$ during slot $t$ across files. We model the temporal evolution of this vector with Markov dynamics as $\pmb r_n (t+1,\tau+1) = \lfloor \pmb r_n (t,\tau+1) + {\pmb \Delta}_n (t,\tau+1)\rfloor^+$, where $\pmb \Delta_n(t,\tau+1)$ is a multivariate Gaussian random noise; $\lfloor \cdot \rfloor$ and $(\cdot)^+$ denote the entry-wise floor and  $\max \left\{\cdot, 0\right\}$ operators. In this context, the incurred cost $\pmb c_0 ({\pmb s}_0(\tau), \pi_0({\pmb s}_0(\tau))) \in {\mathcal C}$ (c.f. \eqref{eq:c0}), is a random vector with an unknown distribution. As requests $\{ \pmb r_n\}$ are random, caching decisions $\{\pmb a_n\}$ are also random. In addition, decision $\pmb a_0$ of the parent node  influences those of leaf nodes via \eqref{eq:leaf_node}, as well as the next state of the parent node in \eqref{eq:netstate}. That is, the current decision of parent node probabilistically influences its next state. To formalize this, we use $P^{\pmb a}_{\pmb s \pmb s'} \in {\mathcal P}$ to denote the \textit{unknown} state transition probability from state $\pmb s$ to $\pmb s'$ upon taking action $\pmb a$
\begin{equation}
\label{eq:transition_prob}
P^{\pmb a}_{\pmb s \pmb s'} = {\textrm {Pr}} \big\{\pmb s(\tau + 1) = \pmb s' \mid \pmb s(\tau) = \pmb s,\, \pmb a = \pi(\pmb s) \big\} 
\end{equation} 
where the subscript $0$ referring to the parent node is dropped for brevity. 
As $\pmb a_0$ probabilistically influences the next state as well as the incurred cost, 
our problem is indeed an MDP. The rest of this paper targets finding an optimal policy solving \eqref{eq:pi00}.} 

{\color{black} \subsection{RL based caching} \label{sec:RL}}
 Towards developing an RL solver for  \eqref{eq:pi00}, define the so-termed \emph{value function} to indicate the quality of policy $\pi_0$, starting from initial state $\pmb s_0 (0)$ as 
	\begin{equation}
	\label{eq:Vdef}
	V_{\pi_0} ( \pmb s_0 ( 0 )):= \mathbb {E} \Big[ \sum_{\tau= 1}^{\infty} \gamma^{\tau-1} {\bf 1}\!^{\top} \pmb \! \pmb c_0 \! \left( \pmb s_0 (\tau), \pi_0 ( \pmb s_0 \left(\tau \right) ) \right)  \Big]
	\end{equation}  
which represents the average cost of following policy~$\pi_0$ to make caching decisions starting from $\pmb s_0(0)$. Upon finding $V_{\pi_0}(\cdot)$ for all $\pi_0 \in \Pi_0$, one can readily obtain $\pi_0^\ast$ that minimizes $V_{\pi_0}(\pmb s_0 ( 0 ))$ over all possible initial states $ \pmb s_0 (0)$.
	
For brevity, the time index and the subscript $0$ referring to the parent node will be dropped whenever it is clear from the context.
To find $V_\pi(\cdot)$, one can rely on the Bellman equation, which basically relates the value of a policy at one state to values of the remaining states \cite[p.~46]{RLbook}. 
Leveraging \eqref{eq:Vdef} and \eqref{eq:transition_prob}, the Bellman equation for value function $V_\pi(\cdot)$ is given by
\begin{equation} \label{eq:VBellman} 
	V_{\pi} ( \pmb s) =  \mathbb {E} \Big[{\bf 1}\!^{\top} \! {\pmb c} (\pmb s, \pi (\pmb s ))  + \gamma \sum_{\pmb s'} P^{\pi (\pmb s)}_{\pmb s \pmb s'} V_{\pi} (\pmb s')  \Big]
	\end{equation}
	where the average immediate cost can be found as 
	\begin{equation*}
	\mathbb{E}\! \left[ {\bf 1}\!^{\top} \!{\pmb c} (\pmb s, \pi (\pmb s )) \right] = \sum_{\pmb s'} P^{\pi (\pmb s)}_{\pmb s \pmb s'} {\bf 1}\!^{\top}\! {\pmb c}\! \left(\pmb s, \pi (\pmb s) | \pmb s'\right). 
	\end{equation*}

If $P^{\pmb a}_{\pmb s \pmb s'}$ were known $\forall \pmb a \in {\mathcal A}$, $\forall {\pmb s, \pmb s' \in {\mathcal S}}$, finding $V_{\pi}(\cdot)$ would be equivalent to solving the system of linear equations \eqref{eq:VBellman}. Indeed, if one could afford the complexity of evaluating $V_{\pi} ( \cdot )$ for all $\pi\in\Pi_0$, the optimal policy $\pi^\ast$ is the one that minimizes the value function for all states. However, the large
(possibly infinite)  number of policies in practice discourages such an approach. An alternative is to employ the so-termed policy iteration algorithm \cite[p.~64]{RLbook} outlined next. Define first the state-action value function, also called $Q$-function for policy~$\pi$ 
\begin{equation}
	\label{eq:Qdef}
	Q_{\pi} \! \left(\pmb s, \pmb a\right) := {\mathbb{E}}\! \left[ {\bf 1}\!^{\top}\! {\pmb c} \left(\pmb s, \pmb a \right)  \right] + \gamma \sum_{\pmb s'} P^{\pmb a}_{\pmb s \pmb s'} V_{\pi}\! \left(\pmb s'\right).
\end{equation}
This function captures the expected immediate cost of starting from state $\pmb s$, taking the first action to be $\pmb a$, and subsequently following policy $\pi$ to take future actions onwards. 
The only difference between the value function in~\eqref{eq:Vdef} and that of $Q$-function in \eqref{eq:Qdef} is that the former takes the first action according to policy $\pi$, while the latter starts with $\pmb a$ as the first action, which may not necessarily be taken when adhering to~$\pi$. Having defined the $Q$-function, we are ready to present the policy iteration algorithm, in which every iteration $i$ entails the following two steps:
	
\emph{Policy Evaluation.} Find $V_{\pi^i} (\cdot)$ for the current policy ${\pi^i}$ by solving the system of linear equations in \eqref{eq:VBellman}.
	
\emph{Policy Improvement.} Update the current policy greedily as
	\begin{equation*}
	{\pi^{i+1}} (\pmb s) = \arg \min_{\pmb \alpha \in \mathcal A} \;Q_{\pi^i} (\pmb s, \pmb \alpha).
	\end{equation*}
	
To perform policy evaluation, we rely on knowledge of  $P^{\pi^i\!(\pmb s)}_{\pmb s \pmb s'}$. However, this is impractical in our setup that involves dynamic evolution of file requests, and unknown caching policies for the leaf nodes. This calls for approaches that target directly the optimal policy $\pi^\ast$, without knowing $P^{\pmb a}_{\pmb s \pmb s'}, \,\forall \pmb a,\pmb s,\, \pmb s'$. One such approach is $Q$-learning~\cite[p.~107]{RLbook}. 

In the ensuing section, we first introduce a $Q$-learning based adaptive caching scheme, which is subsequently generalized in Section~\ref{sec:DQN} by invoking a DQN. 
	
\subsection{Q-learning based Adaptive Caching}
The $Q$-learning algorithm finds the optimal policy $\pi^{\ast}$ by estimating $Q_{\pi^{\ast}}(\cdot,\cdot)$ `on-the-fly.' It turns out that $\pi^{\ast}$ is the greedy policy over the optimal  $Q$-function~\cite[p. 64]{RLbook}, that is 	
\begin{equation}
	\pi^{\ast} (\pmb s) = \arg \min_{\pmb \alpha \in \mathcal A}\, Q_{\pi^{\ast}} (\pmb s, \pmb \alpha ), \quad \forall \pmb s \in {\mathcal S}\label{eq:pistar}
\end{equation}
where $Q_{\pi^\ast}$ is estimated using Bellman's equation for the $Q$-function. This is possible because the $V$-function is linked with the $Q$-function under $\pi^\ast$ through (see~\cite[p.~51]{RLbook} for details)
\begin{equation}
	\label{eq:VstarQstar}
	V_{\pi^{\ast}} (\pmb s) =  \min_{\pmb \alpha \in {\mathcal A}}\; Q_{\pi^{\ast}} (\pmb s, \pmb \alpha ), \quad  \forall \pmb s.
\end{equation}
Substituting \eqref{eq:VstarQstar} into~\eqref{eq:Qdef}, Bellman's equation for  the $Q$-function under $\pi^\ast$ is expressed as 
\begin{equation}
	\label{eq:QBellman}
	Q_{\pi^{\ast}} (\pmb s, \pmb a) = {\mathbb{E}} \!\left[ {\bf 1}^{\top}\!{\pmb c}(\pmb s,\pmb a)  \right] +\gamma \sum_{\pmb s'} P^{\pmb a}_{\pmb s \pmb s'} \min_{\pmb \alpha}\, Q_{\pi^{\ast}}\! (\pmb s', \pmb \alpha )
\end{equation}
which plays a key role in many RL algorithms. Examples include $Q$-learning~\cite{Qlearning}, and SARSA~\cite{RLbook}, where one relies on \eqref{eq:QBellman} to update estimates of the $Q$-function in a stochastic manner. 
In particular, the $Q$-learning algorithm follows an exploration-exploitation procedure to take some action $\pmb a$ in a given state $\pmb s$. Specifically, it chooses the action minimizing the current estimate of $ Q_{\pi^{\ast}} (\cdot,\cdot)$ denoted by $\hat Q_{\tau} (\cdot,\cdot)$, with probability (w.p.) $1-\epsilon_\tau$, or, it takes a random action $\pmb a\in\mathcal{A}$ otherwise; that is,
\begin{equation*}
	\pmb a=\!\left\{\! \!\!\begin{array}{ll}
	\underset{\pmb \alpha \in \mathcal A}{\arg \min} \;~\hat Q_{\tau} (\pmb s,\pmb \alpha ),&{\rm {w.p.} \; 1- \epsilon_\tau}\\	{\rm { random }~\;\pmb a \in {\mathcal A}},&{\rm {w.p.} \;\; \epsilon_\tau}
\end{array}
	\right..
	\end{equation*}
After taking action $\pmb a$, moving to some new state $\pmb s'$, and incurring cost $\pmb c$, the $Q$-learning algorithm adopts the following loss function for the state-action pair $(\pmb s,\pmb a)$ 
\begin{equation}
	{\mathcal L}(\pmb s, \pmb a) =  \frac{1}{2} 
	 \Big( {\bf 1}\!^{\top} {\pmb c}(\pmb s, \pmb a) + \gamma \min_{\pmb \alpha \in \mathcal A} \,\hat Q_\tau (\pmb s' ,\pmb \alpha ) - \hat Q_\tau ( \pmb s , \pmb a) \Big)^2\!\!.
	\label{eq:losssa}
\end{equation}
The estimated $Q$-function for a \textit{single} state-action pair is subsequently updated, by following a gradient descent step to minimize the loss  in \eqref{eq:losssa}, which yields the update
\begin{equation}
	\label{eq:qhatsa}
	\hat Q_{\tau+1} (\pmb s, \pmb a)  = \hat Q_{\tau} (\pmb s, \pmb a )  - \beta \frac{\partial {\mathcal L}(\pmb s, \pmb a )}{\partial \hat Q_{\tau} (\pmb s, \pmb a )}
\end{equation}
where $\beta>0$ is some step size. Upon evaluating the gradient and merging terms, the update in \eqref{eq:qhatsa} boils down to 
\begin{align*}
	\hat Q_{\tau+1} (\pmb s, \pmb a ) \!=\!(1 \!-\! \beta) \hat Q_\tau (\pmb s, \pmb a )\! +\! \beta \Big[ {\bf 1}\!^{\top} \!\! {\pmb c} ( \pmb s, \pmb a )\! +\!\gamma \min_{\pmb \alpha \in \mathcal A}\hat Q_\tau (\pmb s' ,\pmb \alpha)  \Big].
\end{align*}	

Three remarks are worth making at this point.
\begin{remark}
As far as the fast-timescale caching strategy of leaf nodes is concerned, multiple choices are possible, including e.g., LRU, LFU, FIFO,~\cite{fagin1977asymptotic}, the multi-armed bandit scheme~\cite{multiarm2014}, and even RL ones~\cite{RL1, RL2}.
\end{remark}
		
\begin{remark}	
The exploration-exploitation step for taking actions guarantees continuously visiting state-action pairs, and ensures convergence to the optimal $Q$-function~\cite{Tsitsiklis}. Instead of the $\epsilon$-greedy exploration-exploitation step, one can employ the upper confidence bound scheme~\cite{Conf_expl}. Technically, any exploration-exploitation scheme should be greedy in the limit of infinite exploration (GLIE)~\cite[p.~840]{2016artificial}. An example obeying GLIE is the $\epsilon$-greedy algorithm \cite[p.~840]{2016artificial} with $\epsilon_\tau = 1/\tau$. It converges to an optimal policy, albeit at a very slow rate. On the other hand, using a constant $\epsilon_\tau=\epsilon$ approaches the optimal $Q^\ast(\cdot, \cdot)$ faster, but its exact convergence is not guaranteed as it is not GLIE.
\end{remark}

Clearly, finding $Q_{\pi^\ast} (\pmb s, \pmb a)$ entails estimating a function defined over state and action spaces. In several applications however, at least one of the two vector variables is either continuous or takes values from an alphabet of high cardinality. Revisiting every state-action pair in such settings is impossible, due to the so-called curse of dimensionality -- a typical case in practice. To deal with it, function approximation techniques offer as a promising solution~\cite{RLbook}. These aim at finding the original $Q$-function over all feasible state-action pairs, by judicious generalization from a few observed pairs. 
Early function approximators for RL design good hand-crafted features that can properly approximate $V_{\pi^\ast}(\cdot)$, $Q_{\pi^\ast}(\cdot,\cdot)$, or, $\pi^\ast(\cdot)$~\cite{minh2015}. Their applicability has only been limited to domains, where such features can be discovered, or, to state spaces that are low dimensional~\cite{RLbook}.

\begin{figure}
	\centering
	\includegraphics[width = 0.48 \textwidth]{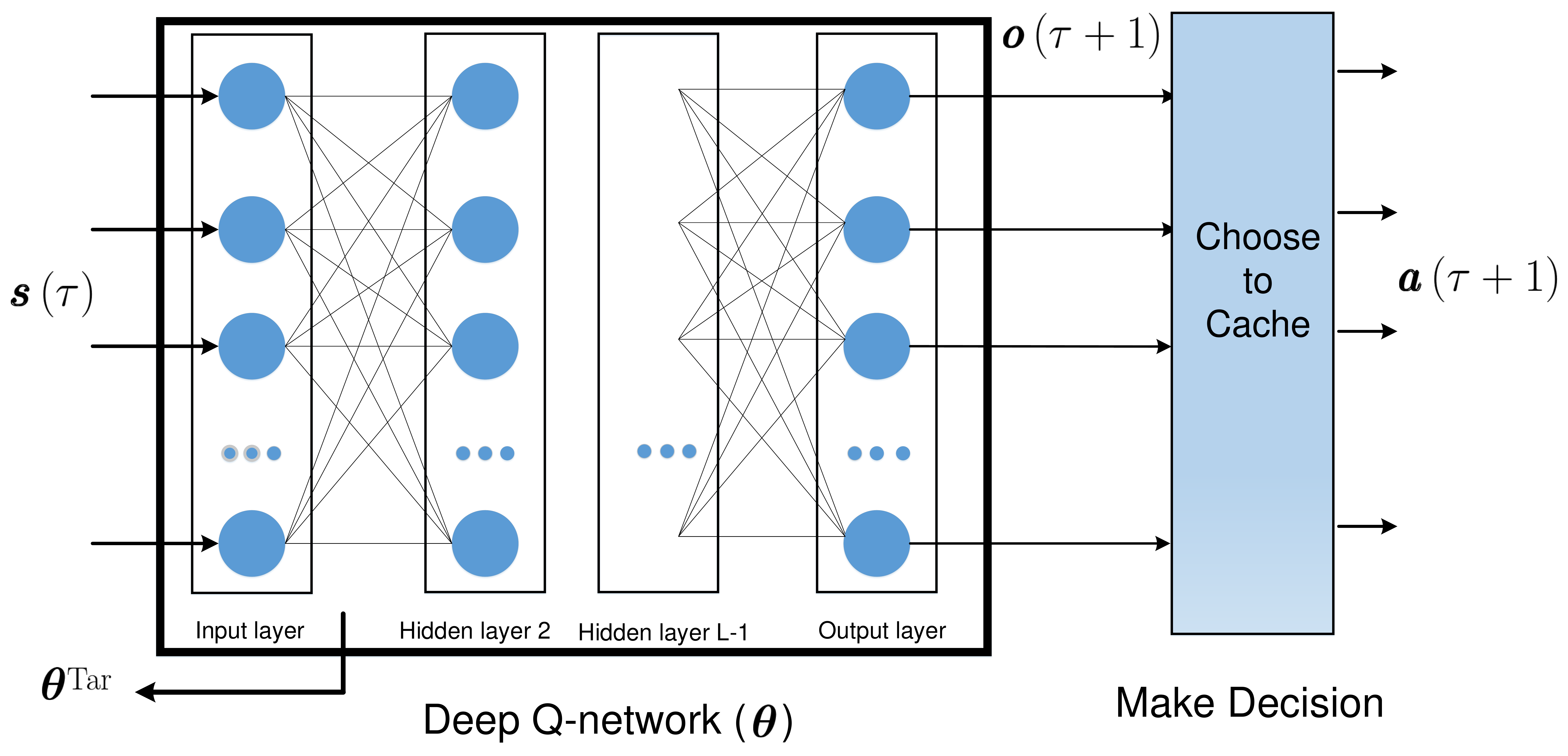}
	\caption{Deep $Q$-network}
	\label{fig:DQN}
\end{figure}

Deep learning approaches on the other hand, have recently demonstrated remarkable potential in applications such as object detection, speech recognition, and language translation, to name a few~\cite{Briefsurveydrl}. This is because DNNs are capable of extracting compact low-dimensional features from high-dimensional data. Wedding DNNs with RL results in `deep RL' that can effectively deal with the curse of dimensionality, by eliminating the need of hand-crafting good features. These considerations have inspired the use of DNNs to estimate either the $Q$-function, value function, or, the policy~\cite{Briefsurveydrl}. The most remarkable success in playing AI games adopted DNNs to estimate the $Q$-function, what  is termed DQN in~\cite{minh2015}. 

Prompted by this success, the next leverages the power of function approximation through DNNs to develop an adaptive caching scheme for the parent node. 

\section{Adaptive DQN-Based Caching} \label{sec:DQN}
To find the optimal caching policy $\pi^\ast$ for the parent node, the success of function approximators for RL (e.g., \cite{minh2015}), motivated us to pursue a parametric approach to estimating $Q(\pmb s,  \pmb a)$ with a DNN (cf. \eqref{eq:pistar}). This DNN has as input  pairs of vectors $(\pmb s, \pmb a)$, and as scalar output the corresponding estimated $Q(\pmb s, \pmb a)$ values. Clearly, the joint state-action space of the sought Q-function has cardinality $|{\mathcal A} \times \mathcal S| = |{\mathcal A}| |\mathcal S|$. To reduce the search space, we shall take a parametric DQN approach \cite{minh2015} that we adapt to our setup, as elaborated next. 
		
Consider a deep feedforward NN with $L$ fully connected layers, with input the $F\times 1$ state vector $\pmb s(\tau)$ as in \eqref{eq:netstate}, and $F\times 1$ output cost vector  $\pmb o (\tau + 1)$~\cite{minh2015}. Note that input does not include the action vector $\pmb a$. Each hidden layer $l \in \{ 2, \ldots, L-1\}$ comprises $n_l$ neurons with rectified linear unit (ReLU) activation functions $h(z):=\max(0,\,z)$ for $z \in {\mathbb R}$; see e.g., \cite{relus}. Neurons of the $L$-th output layer use a softmax nonlinearity \footnote{Softmax is a function that takes as input a vector $\bm{z}\in\mathbb{R}^F$, and normalizes it into a probability distribution via $\sigma({\bm z})_f={\rm e}^{{ z}_f}/(\sum_{f=1}^F {\rm e}^{z_f})$, $\forall f$.} to yield for the $f$-th entry of $\pmb o$, an estimated long term cost $o_{f}$ that would incur, if file $f$ is not stored at the cache.

If this cache memory can store up to $M$ ($<F$) files at the parent node, the $M$ largest entries of the DQN output $\pmb o (\tau + 1)$ are chosen by the decision module in Fig.~\ref{fig:DQN} to obtain the action vector $\pmb a (\tau + 1)$. We can think of our DQN as a `soft policy' function estimator, and $\pmb o (\tau + 1)$ as a `predicted cost' or a `soft action' vector for interval $\tau+1$, whose `hard thresholding' yields $\pmb a (\tau + 1)$.  It will turn out that excluding $\pmb a$ from the DQN input and picking it up at the output lowers the search space from 
$|{\mathcal A}| |\mathcal S|$ to $|\mathcal S|$. 
		
To train our reduced-complexity DQN amounts to finding a set of weights collected in the vector~$\boldsymbol \theta$ that parameterizes the input-output relationship $\pmb o (\tau +1 ) = Q(\pmb s (\tau); \boldsymbol \theta_\tau)$. To recursively update 
$\boldsymbol \theta_\tau$ to ${\boldsymbol \theta}_{\tau+1}$, consider two successive intervals along with corresponding states $\pmb s(\tau)$ and $\pmb s(\tau+1)$; the action  $\pmb a(\tau+1)$ taken at the beginning of interval $\tau+1$; and, the cost $\pmb c({\tau}+1)$ revealed at the end of interval $\tau +1$. The instantaneous approximation of the optimal cost-to-go from interval $\tau +1$ is given by $\pmb c({\tau}+1) + \gamma Q(\pmb s(\tau+1) ; \pmb \theta_\tau)$, where $\pmb c({\tau}+1)$ is the immediate cost, 
and $ Q(\pmb s(\tau+1) ; \pmb \theta_\tau)$ represents the predicted cost-to-go from interval $\tau +2$ that is provided by our DQN with $\boldsymbol \theta_\tau$, and discounted by $\gamma$. Since our DQN offers  $Q (\pmb s({\tau}) ; \pmb \theta_{\tau})$ as the predicted cost for interval $\tau+1$, 
the prediction error of this cost as a function of  $\boldsymbol \theta_\tau$ is given by
%
\begin{align}
\nonumber
{\pmb \delta}(\pmb \theta_\tau )  &\,:=  [ \overbrace{\pmb c({\tau}+1) + \gamma Q(\pmb s(\tau+1) ; \pmb \theta_\tau)}^{ \textrm {target cost-to-go from interval $\tau+1$}} - Q (\pmb s({\tau}) ; \pmb \theta_{\tau}) ]  \\ 	& \quad ~~\odot ({\bf 1} - {\pmb a} (\tau +1 ) )
\label{eq:errorDQN}
\end{align}
%
and has non-zero entries for files not stored at interval $\tau+1$.

Using the so-termed experience ${\pmb  E}_{\tau+1} := [\pmb s(\tau),   \pmb a(\tau+1), \pmb c(\tau+1), \pmb s(\tau + 1)]$, and 
the $\ell_2$-norm of ${\pmb \delta}(\pmb \theta_\tau)$ as criterion 
\begin{equation}
		\label{eq:lossDQN}
		{\mathcal L}  (\pmb \theta_{\tau}) = \big\|{\pmb \delta} (\pmb \theta_\tau ) \big\|_2^2
\end{equation}
the sought parameter update minimizing \eqref{eq:lossDQN} is given by the stochastic gradient descent (SGD) iteration as   
\begin{equation}
		\label{eq:SGD}
		\boldsymbol \theta_{\tau+1} = \boldsymbol \theta_{\tau} - \beta_\tau \!\! \left. \nabla \! {\mathcal L} (\boldsymbol \theta) \right |_{\boldsymbol \theta = \bm \theta_{\tau}}
\end{equation}   
where $\beta_{\tau}>0$ denotes the learning rate.  
		
Since the dimensionality of $\pmb \theta$ can be much smaller than $|\mathcal S|  |{\mathcal A}|$, the DQN is efficiently trained with few experiences, and generalizes to unseen state vectors. Unfortunately,  DQN model inaccuracy can propagate in the cost prediction error in \eqref{eq:errorDQN} that can cause instability in \eqref{eq:SGD}, which can lead to performance degradation, and even divergence~\cite{silver2016,Machine2016}. Moreover, \eqref{eq:SGD} leverages solely a single most recent experience ${\pmb  E}_{\tau+1}$. These limitations will be mitigated as elaborated next. 
		
\begin{algorithm}[t]
			\SetKwInOut{Input}{Initialize}
			\SetKwInOut{Output}{Output}
			\Input{$\pmb s (0)$,  $\pmb s_n (t,\tau), \forall n$, $\pmb \theta_{\tau}$, and $\pmb \theta^{\textrm{Tar}}$ }
			\For{$\tau = 1, 2, \ldots $}
			{ Take action $\pmb a(\tau)$ via exploration-exploitation 
				\\
				{ {$\pmb a (\tau) = \bigg\{ \begin{matrix}
						{\rm{Best\; files\; via}} \; Q(\pmb s (\tau\!-\!1);\pmb \theta_{\tau}) \hfill {~\rm {w.p.} \; 1- \epsilon_\tau}
						\\	{\rm { random } \; \pmb a \in {\mathcal A}}  \hfill {\rm {w.p.} \; \epsilon_\tau} \end{matrix} $} 
					\\
					\For{$t = 1, \ldots , T$}
					{\For{$n = 1, \ldots, N$}{Take action $\pmb a_{n}$ using local policy 
							\\ 
							$\pmb a_n (t, \tau) = \Big\{ \begin{matrix} \!\pi_n (\pmb s ( t -1, \tau )) \; {\rm if} \; t \ne 1
							\\ 
							\pi_n (\pmb s ( T , \tau -1 )) \; {\rm if} \; t = 1 
							\end{matrix} $ 
							\\ 
							{Requests $\pmb r_{n} (t,\tau)$ are revealed 				
								\\ 
								{Set $\pmb s_n (t,\tau) = \pmb r_n (t,\tau)$}
								\\ 
								{Incur $\pmb c_n (\cdot)$}, cf. \eqref{eq:nodalcost}}     	 	
						}}
						{\bf {Leaf nodes}} 
						\\
						{Set  \hspace{+.4 cm}${\bar {\pmb s}_n} (\tau):= (1/T) \sum_{t=1}^{T} \; \pmb s_{n} ( t, \tau )$} 
						\\ 	
						{Send \hspace{+.05 cm} ${\bar {\pmb s}_{n}} ( \tau ) \odot ({\bf 1} - \pi_n ( {\bar {\pmb s}_{n}} ) )$ to parent node}
						\\ 
						{Send \hspace{+.2 cm}${\bar {\pmb c}}_n (\cdot)$} cf. \eqref{eq:cost_pernode}, to parent node 
						\\ 
						{\bf {Parent node}} 
						\\ 
						Set \hspace{+.3 cm} $\pmb s (\tau):= \sum_{n = 1}^{N} w_n  {\bar {\pmb s}_{n}} ( \tau ) \odot ({\bf 1} - \pi_n ( {\bar {\pmb s}_{n}} ) )$ \\ 
						Find \,\,   $\pmb c (s({\tau}\!-\!1), \pmb a({\tau}))\;$ 
						\\
						{{Save \hspace{+.00 cm} $(\pmb s({\tau}\!-\!1), \pmb a({\tau}), \pmb c(s({\tau}\!-\!1), \pmb a({\tau})), \pmb s({\tau}) )$} in $\mathcal E$} 
						\\
						{Uniformly sample $B$ experiences from $\mathcal E$} 
						\\
						{Find $  \nabla {\mathcal L}^{\textrm{Tar}} (\boldsymbol \theta) $ for these samples, using \eqref{eq:lossTar} }
						\\
						{Update~$\boldsymbol \theta_{\tau+1} = \boldsymbol \theta_{\tau} - \beta_\tau \nabla {\mathcal L}^{\textrm{Tar}} (\boldsymbol \theta)$} 
						\\
						{If $\rm mod (\tau,C) = 0$, then update $\pmb \theta^{\textrm{Tar}} = {\pmb \theta}_\tau$}	
					} }
					\caption{Deep RL for adaptive caching.}
					\label{Alg_2}
				\end{algorithm}

\subsection{Target Network and Experience Replay}
NN function approximation, along with the loss~\eqref{eq:lossDQN} and the update~\eqref{eq:SGD}, often result in unstable RL algorithms~\cite{minh2015, lillicrap2016}. This is due to: i) correlated experiences used to update the DQN parameters $\pmb \theta$; and, ii) the influence of any change in policy on subsequent experiences and vice versa. 
				
Possible remedies include the so-called {\it experience replay} and {\it target network} to update the DQN weights. In experience replay, the parent node stores all past experiences ${\pmb  E}_{\tau}$ in ${\mathcal E}:= \{\pmb E_{1}, \ldots, \pmb E_{\tau} \}$, and utilizes a batch of $B$ uniformly sampled experiences from this data set, namely $\{\pmb E_{i_\tau}\}_{i= 1}^{B} \sim U (\mathcal E)$. By sampling and replaying previously observed experiences, experience replay can overcome the two challenges. On the other hand, to obtain decorrelated target values in \eqref{eq:errorDQN}, a second NN (called target network) with structure identical to the DQN is invoked with parameter vector  $\boldsymbol \theta^{\textrm {Tar}}$. Interestingly, $\boldsymbol \theta^{\textrm {Tar}}$ can be periodically replaced with $\boldsymbol \theta_{\tau}$ every $C$ training iterations of the DQN, which enables the target network to smooth out fluctuations in updating the DQN~\cite{minh2015}. 
				
With a randomly sampled experience $\pmb E_{i_\tau}\in\mathcal{E}$, 
the prediction error with the target cost-to-go estimated using the target network (instead of the DQN) is 
\begin{align}
				\nonumber
				{\pmb \delta}^{\textrm{Tar}}(\pmb \theta ;\pmb E_{i_\tau}) &:= \!    [{\pmb c(i_{\tau}\! +\! 1) + \!  \gamma  Q( {\pmb s(i_{\tau}\! +\! 1)}; {\boldsymbol \theta}^{\textrm {Tar}} )}- Q(\pmb s({i_\tau}) ;\boldsymbol \theta )] \\  & ~\quad \odot ({\bf 1} - \pmb a({i_\tau\! +\! 1}) ). \label{eq:errorTarget}
\end{align}
Different from \eqref{eq:errorDQN}, the target values here are found through the target network with weights $\pmb \theta^{\textrm{Tar}}$. In addition, the error in \eqref{eq:errorDQN} is found by using the most recent experience, while the experience here is randomly drawn from past experiences in $\mathcal E$. As a result, the loss function becomes
\begin{equation}
				\label{eq:lossTar}
				{\mathcal L}^{\textrm{Tar}} (\pmb \theta) = {\mathbb E} \| {\pmb \delta}^{\textrm{Tar}}(\pmb \theta; \pmb E ) \|_2^2
\end{equation}
where the expectation is taken with respect to the uniformly sampled experience $\pmb E$. In practice however, only a batch of $B$ samples is available and used to update $\pmb \theta_{\tau}$, so the expectation will be replaced with the sample mean. Finally, following a gradient descent step over the sampled experiences, we have 
\begin{equation*}
				\boldsymbol \theta_{\tau+1} = \boldsymbol \theta_{\tau} - \beta_\tau \nabla {\mathcal L}^{\textrm{Tar}}\! \left(\boldsymbol \theta\right)|_{\boldsymbol \theta = \bm \theta_{\tau}}.
\end{equation*}
Both the experience replay and the target network help stabilize the DQN updates.  Incorporating these remedies, Alg.~\ref{Alg_2} tabulates our deep RL based adaptive caching scheme for the parent node. 

\begin{figure}
					\centering\includegraphics[width =0.44 \textwidth]{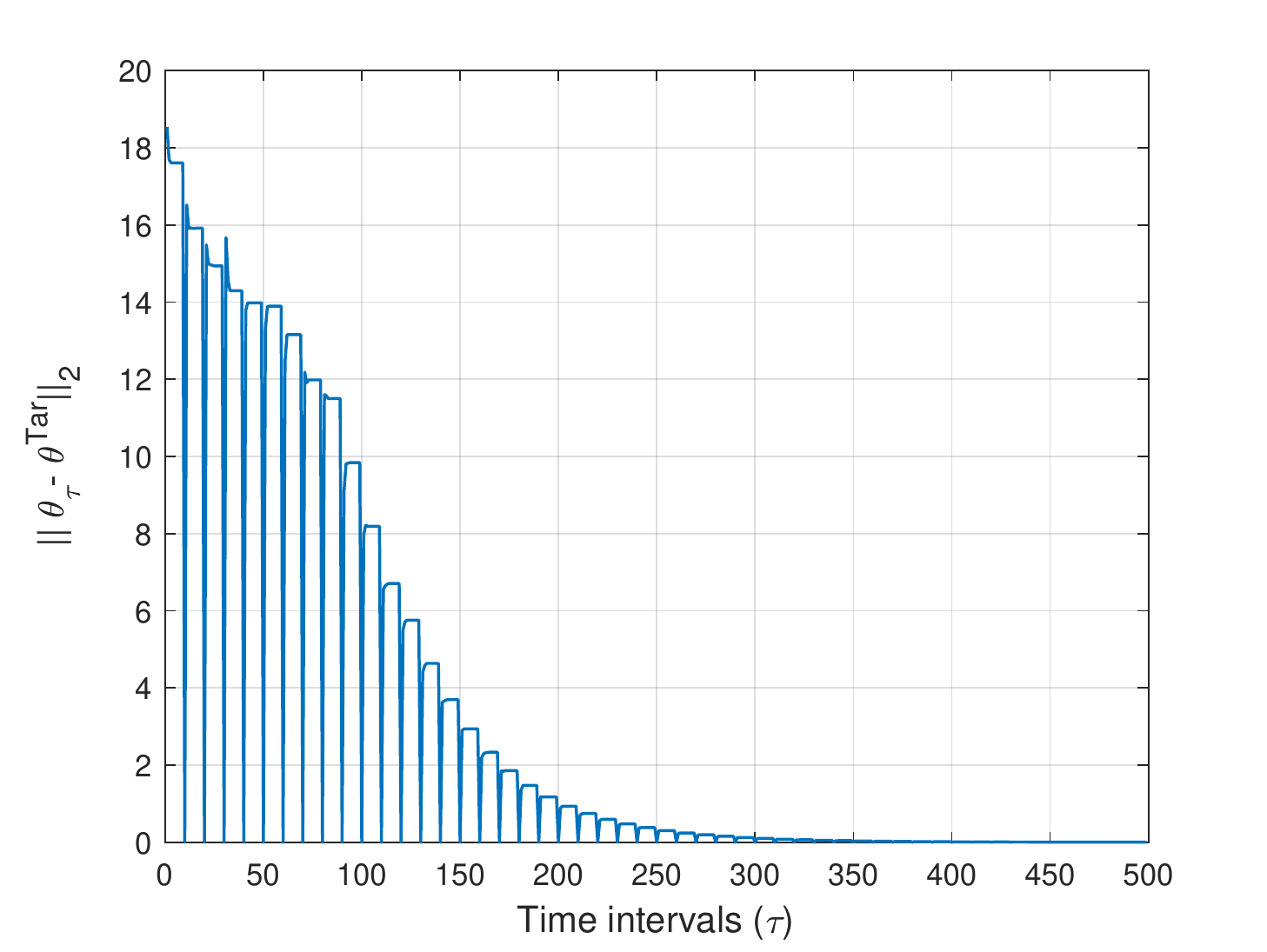}
					\caption{Convergence of DQN to the target network.}
					\label{fig:static_1}
\end{figure}
\begin{figure}
					\centering
					\includegraphics[width =0.46\textwidth]{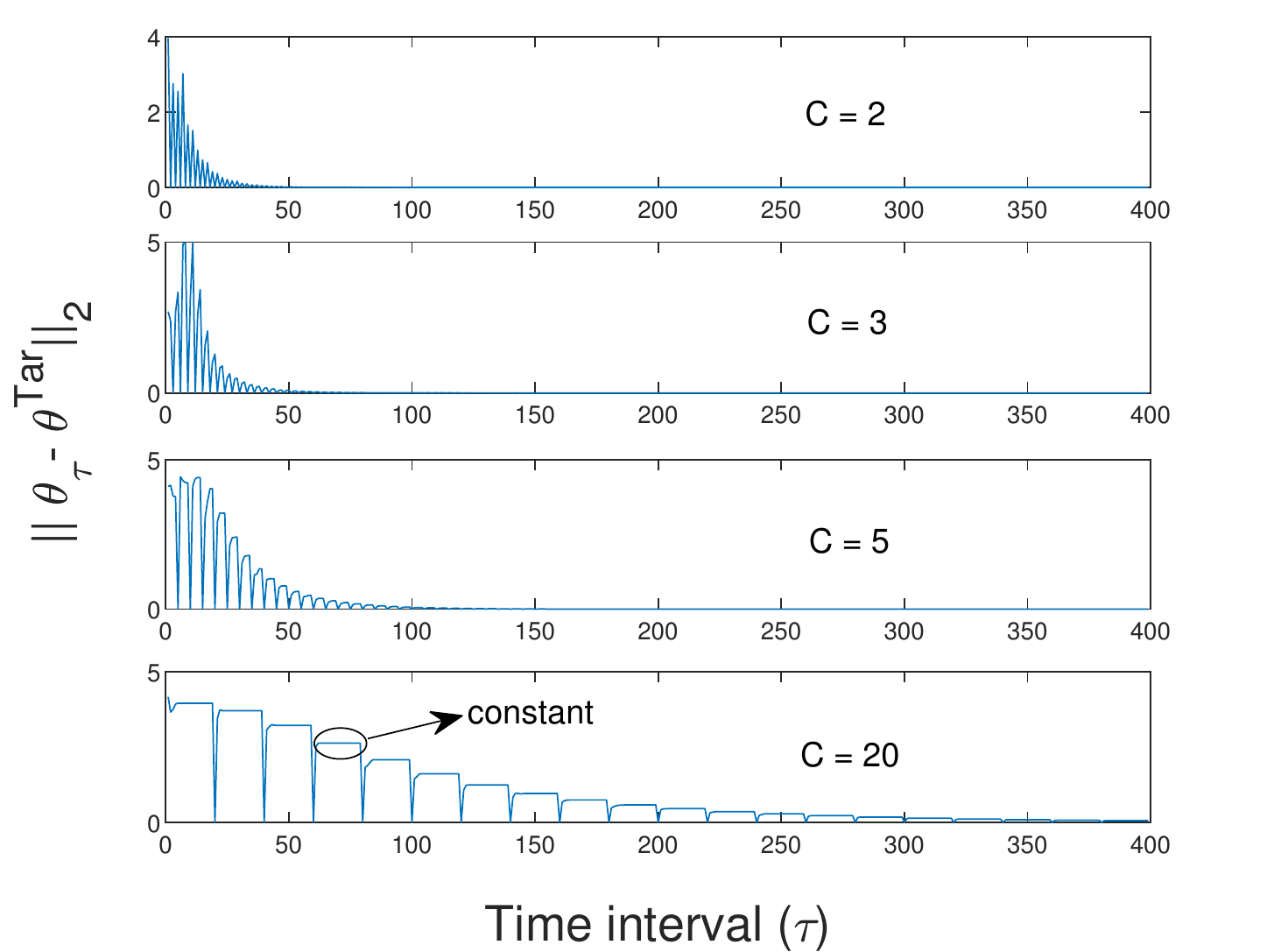}
					\caption{Impact of $C$ on DQN convergence in a static setting.}
					\label{fig:static_2}
\end{figure}
\section{Numerical Tests} 
\label{sec:numerical}
In this section, we present several numerical tests to assess the performance of our deep RL based caching schemes which  is represented in Alg.~\ref{Alg_2}.
				
The first experiment considers a simple setup, consisting of a parent node that directly serves user file requests. A total of $F = 50$ files were assumed in the cloud, each having the same size, while $M_0 = 5$ files can be stored in the cache, amounting to $10\%$ of the total. The file popularities were drawn randomly from $[0,1]$, invariant across all simulated time instants, but unknown to the caching entity. A fully connected feed-forward NN of $3$ layers was implemented for DQN with each layer comprising $50$ hidden neurons. For simplicity, we assume that the dataset $\mathcal E$ can store a number $\mathcal R$ of most recent experiences, which is also known as the replay memory. It was set to ${\mathcal R} = 10$ in our test, along with mini-batch size $B = 1$ to update the target network every $C=10$ iterations. {\color{black} In addition, hyper-parameter values $\gamma = 0.8$, learning rate $\beta_{\tau} = 0.01$, as well as $\epsilon_{\tau} = 0.4$ were used throughout our numerical tests.}
				
Figure~\ref{fig:static_1} shows the convergence of the DQN parameter $\pmb \theta_{\tau}$ to that of the target network $\pmb \theta^{\textrm{Tar}}$. Since $\pmb \theta^{\textrm{Tar}}$ is updated with $\pmb \theta_{\tau}$ per $C$ iteration, the error $||\boldsymbol \theta_{\tau} - \boldsymbol \theta^{\rm{Tar}}||_2$ vanishes periodically. While the error changes just in a few iterations after the $\pmb \theta^{\textrm{Tar}}$ update, it is constant in several iterations before the next $\pmb \theta^{\textrm{Tar}}$ update, suggesting that $\pmb \theta_{\tau}$ is fixed across those iterations. This motivates investigating the impact of $C$ on $\pmb \theta_{\tau}$'s convergence. Figure \ref{fig:static_2} shows the convergence of $\pmb \theta_{\tau}$ to $\pmb \theta^{\textrm{Tar}}$ for $C = 20, \,5,\, 3,\, 2$. Starting with $C = 20$, vector $\pmb \theta_{\tau}$ is not updated in most iterations between two consecutive updates of $\pmb \theta^{\textrm{Tar}}$. Clearly, the smaller $C$ is, the faster the convergence for $\pmb \theta_{\tau}$ is achieved. Indeed, this is a direct result of having static file popularity. Based on our numerical tests, having small $C$ values is preferable in dynamic settings too. 		

For the second experiment, a similar setup having solely the parent node, plus $M_0 =5$ and $F = 50$, was considered. Dynamic file popularities were generated with time evolutions modeled by Markov process described earlier in the paper, and again they are unknown to the caching node.  

\begin{figure}[t!]
	\centering
	\includegraphics[width=.4 \textwidth]{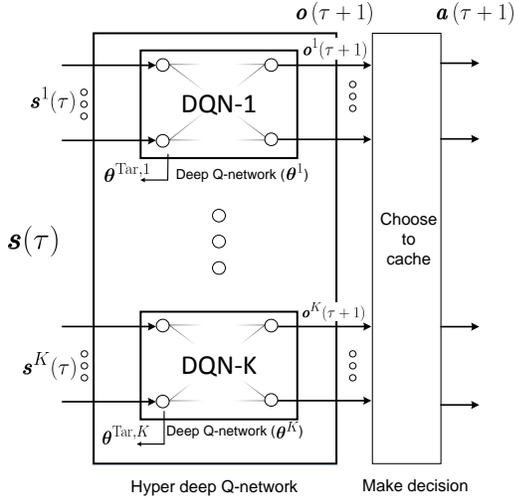}\caption{Hyper deep-Q network for scalable caching.}
	\label{H_DQN}
\end{figure}

{\color{black} We first consider that the parent node is connected to $N = 5$ leaf nodes, and every leaf node implements a local caching policy $\pi_n$ with capacity of storing $M_n = 5$ files. Every leaf node receives user file requests within each fast time slot, and each slow-timescale interval consists of $T = 2$ slots. File popularity exhibits different Markovian dynamics locally at leaf nodes. Having no access to local policies $\{\pi_n\}$, 
the parent node not only should learn file popularities along with their temporal evolutions, but also learn the caching policies of leaf nodes. To endow our approach with scalability to handle $F \gg 1$, we advocate the following hyper Q-network implementation. Files are first split into $K$ smaller groups of sizes $F_1,  \ldots, F_K$ with $F_k\ll F$. This yields the representation $\pmb s^\top(\tau) := [{\pmb s^{1}}^\top\!(\tau ) , \ldots , {\pmb s^{K}}^\top\!(\tau ) ]$, where $\pmb s^k \in {\mathbb R}^{F_k}$. By running $K$ DQNs in parallel, every DQN-$k$ now outputs the associated predicted costs of input files through $\pmb o^k (\tau) \in {\mathbb R}^{F^k}$. Concatenating all these outputs, one obtains the predicted output cost vector of all files as $\pmb o^\top(\tau+1)\!:=\! [  {{\pmb o}^1}^\top\!(\tau+1), \ldots , {{\pmb o}^K}^\top\!(\tau+1)]$; see Sec.~\ref{sec:DQN} for discussion on finding action $\pmb a$ from vector $\pmb o$, and also Fig.~\ref{H_DQN} for an illustration of our hyper Q-network. 

The ensuing numerical tests consider $F = 1,000$ files with $K = 25$ and $\{F_k = 20\}$, where only $M_0 = 50$ can be stored. To further assess the performance, we adopt a \textit{non-causal} optimal policy as a benchmark, which unrealistically assumes knowledge of future requests and stores the most frequently requested files. In fact, this is the best policy that one may ideally implement. Further, practically used cache  schemes including e.g., the LRU, LFU, and FIFO~\cite{dan1990approximate} are also simulated. A difference between LRU, LFU, FIFO, and our proposed approach is that they refresh the cache whenever a request is received, while our scheme refreshes the cache only at the end of every time slot. }

\begin{figure}
	\centering\includegraphics[width =0.45 \textwidth]{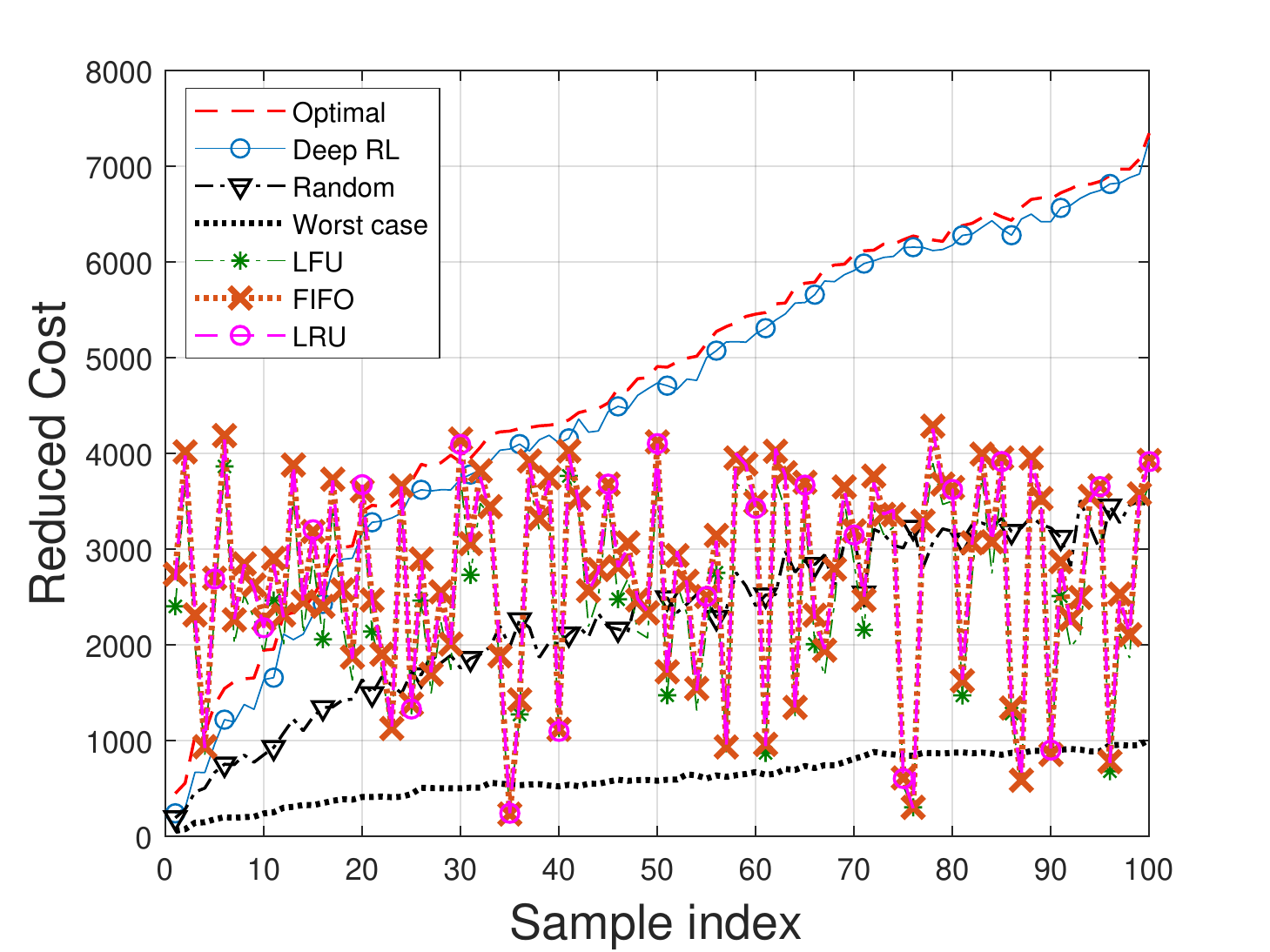}
	\caption{Instantaneous reduced cost for different policies.}
	\label{fig:scalable_1}
\end{figure}

\begin{figure}
	\centering\includegraphics[width =0.48 \textwidth]{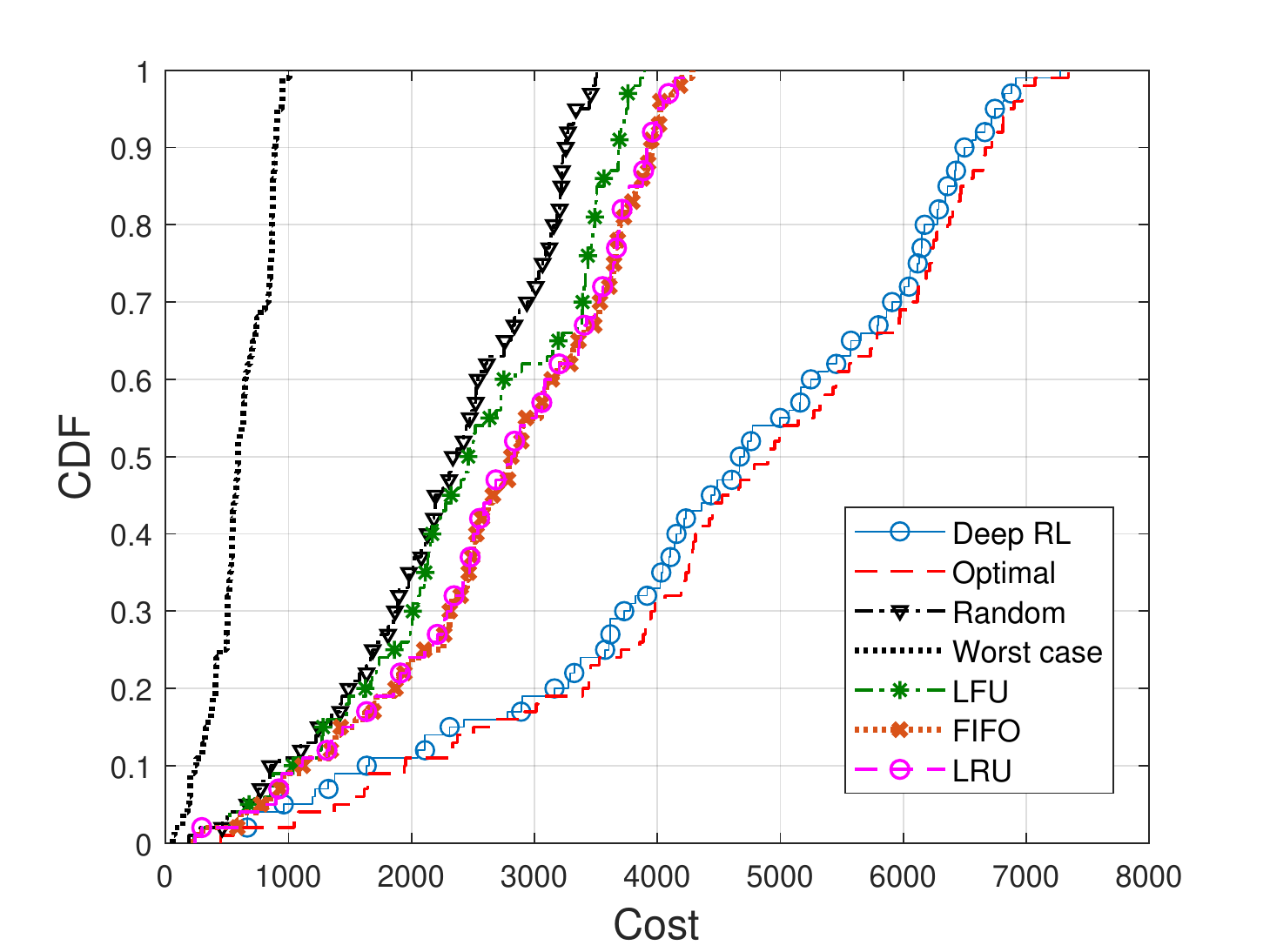}
	\caption{Instantaneous reduced cost for different policies.}
	\label{fig:scalable_2}
\end{figure}

By drawing $100$ samples\footnote{During these samples, DRL policy is enforced to exploit its learned caching policy.} randomly from all $2,000$ time intervals, the instantaneous reduced cost and the empirical cumulative distribution function (CDF) obtained over these $100$ random samples for different policies are plotted in Fig.~\ref{fig:scalable_1} and~Fig.\ref{fig:scalable_2}, respectively. These plots further verify how the DRL policy performs relative to the alternatives, and in particular very close to the optimal policy.  

{\color{black} LRU, LFU, and FIFO make caching decisions based on instantaneous observations, and can refresh the cache many times within each slot. Yet, our proposed policy as well as the optimal one here learns from all historical observations to cache, and refreshes the cache only once per slot. Because of this difference, the former policies outperform the latter at the very beginning of Fig. \ref{fig:scalable_1}, but they do not adapt to the underlying popularity evolution and are outperformed by our learning-based approach after a number of slots. The merit of our approach is further illustrated by the CDF of the reduced cost depicted in Fig.~\ref{fig:scalable_2}.}

\begin{figure}[h!]
	\centering
	{\includegraphics[scale=0.58]{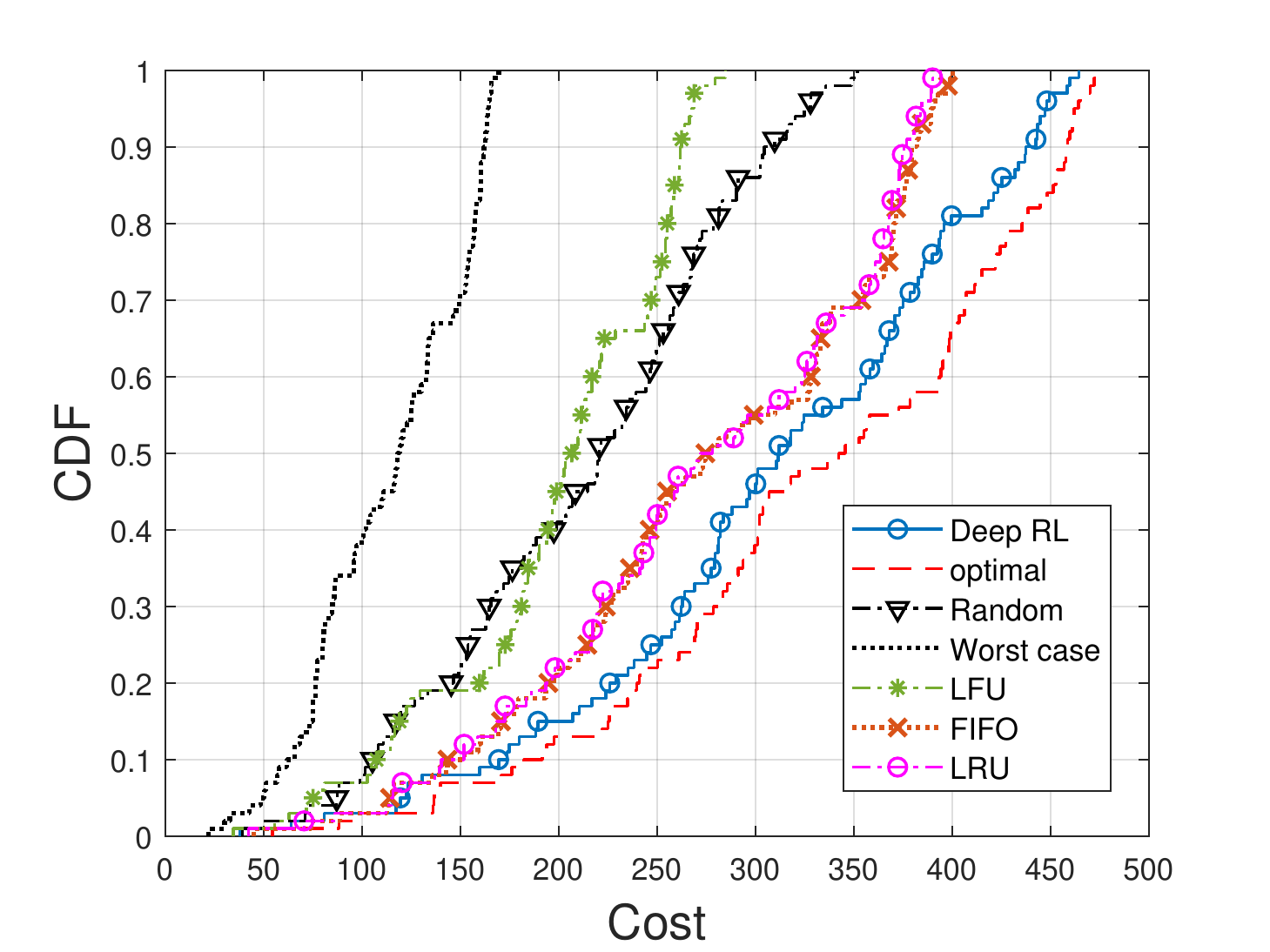}}
	\caption{Performance of all policies for $N = 10$ leaf nodes.}
	\label{fig:Nsubnodes10}
\end{figure}

\begin{figure}[h!]
	\centering
	{\includegraphics[scale=0.58]{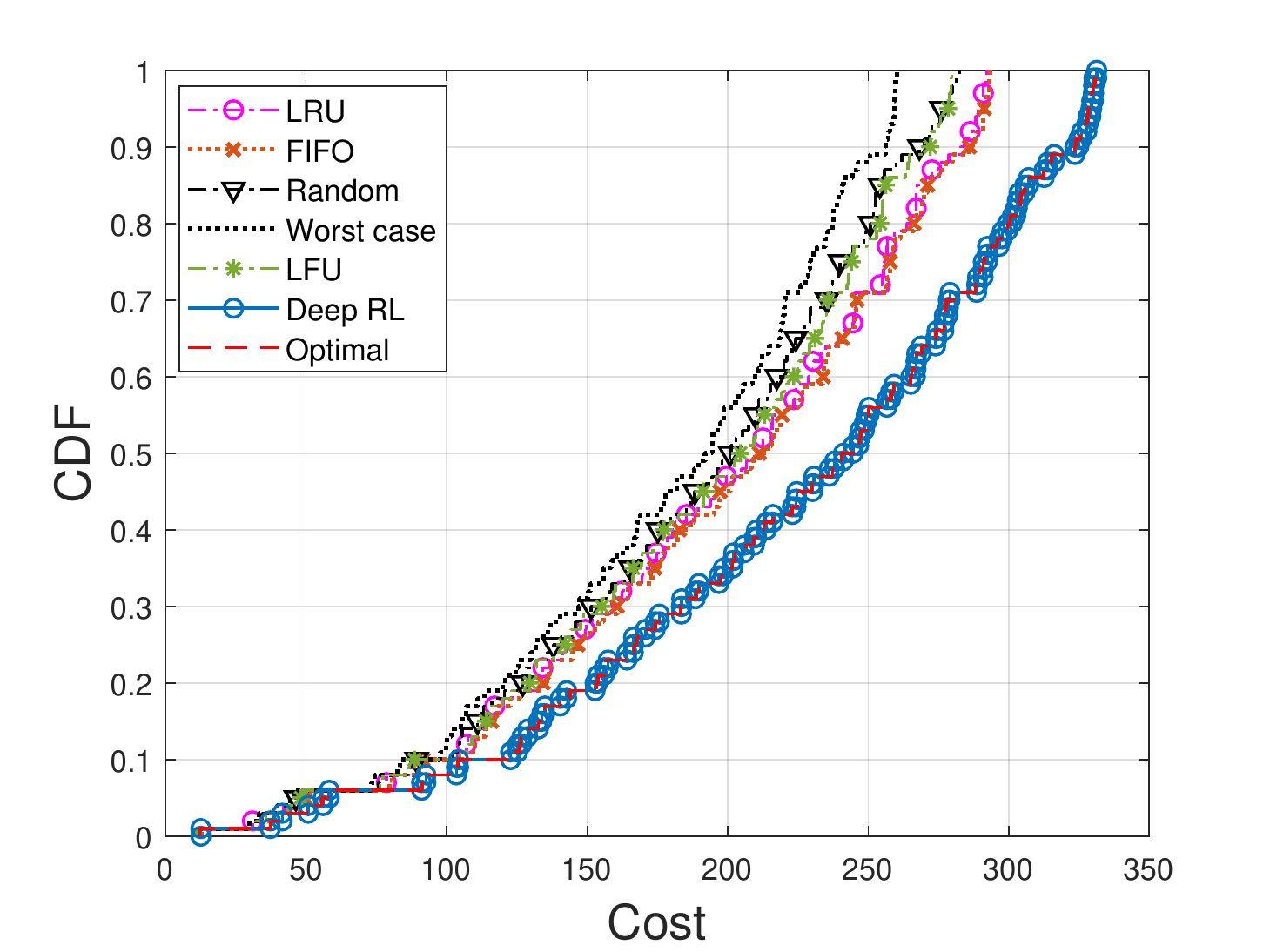}}
	\caption{Performance of all policies for $N = 1,000$ leaf nodes.} 
	\label{fig:Nsubnodes1000}
\end{figure}

In the last test, we increased the number of leaf nodes from $10$ in the previous experiment to~$N = 1,000$. {\color{black} Figures~\ref{fig:Nsubnodes10} and \ref{fig:Nsubnodes1000} showcase that the DRL performance approaches that of the optimal policy as the number of nodes increases. This is likely because the more leaf nodes, the smoother the popularity fluctuations, and therefore the easier it becomes for DRL to learn the optimal policy.}

				
\section{Conclusions}
\label{sec:Conclusion}
Caching highly popular contents across distributed caching entities can substantially reduce the heavy burden of content delivery in modern data networks. This paper considered a network caching setup, comprising a parent node connected to several leaf nodes to serve end user file requests. Since the leaf nodes are closer to end users, they observe file requests in a fast timescale, and could thus serve requests, either through their locally stored files or by fetching files from the parent node. Observing aggregate requests from all leaf nodes, the parent node makes caching decisions in a slower-time scale. Such a two-timescale caching task was tackled in this paper using a RL approach. An efficient caching policy leveraging deep RL was introduced, and shown capable of learning-and-adapting to dynamic evolutions of file requests, and caching policies of leaf nodes. Simulated tests corroborated its impressive performance. 
This work also opens up several directions for future research, including multi-armed bandit online learning \cite{2018lbc}, and distributed deep RL using recurrent NNs \cite{deepcache} for future spatio-temporal file request prediction.

\bibliographystyle{IEEEtran}
\bibliography{biblio}

\begin{thebibliography}{10}
\providecommand{\url}[1]{#1}
\csname url@samestyle\endcsname
\providecommand{\newblock}{\relax}
\providecommand{\bibinfo}[2]{#2}
\providecommand{\BIBentrySTDinterwordspacing}{\spaceskip=0pt\relax}
\providecommand{\BIBentryALTinterwordstretchfactor}{4}
\providecommand{\BIBentryALTinterwordspacing}{\spaceskip=\fontdimen2\font plus
\BIBentryALTinterwordstretchfactor\fontdimen3\font minus
  \fontdimen4\font\relax}
\providecommand{\BIBforeignlanguage}[2]{{%
\expandafter\ifx\csname l@#1\endcsname\relax
\typeout{** WARNING: IEEEtran.bst: No hyphenation pattern has been}%
\typeout{** loaded for the language `#1'. Using the pattern for}%
\typeout{** the default language instead.}%
\else
\language=\csname l@#1\endcsname
\fi
#2}}
\providecommand{\BIBdecl}{\relax}
\BIBdecl

\bibitem{Paschos18}
G.~S. Paschos, G.~Iosifidis, M.~Tao, D.~Towsley, and G.~Caire, ``The role of
  caching in future communication systems and networks,'' \emph{IEEE J. Sel.
  Areas Commun.}, vol.~36, no.~6, pp. 1111--1125, June 2018.

\bibitem{Bastug14}
E.~Bastug, M.~Bennis, and M.~Debbah, ``Living on the edge: {T}he role of
  proactive caching in {5G} wireless networks,'' \emph{IEEE Commun. Mag.},
  vol.~52, no.~8, pp. 82--89, Aug. 2014.

\bibitem{goodfellow2016deep}
I.~Goodfellow, Y.~Bengio, A.~Courville, and Y.~Bengio, \emph{Deep
  {L}earning}.\hskip 1em plus 0.5em minus 0.4em\relax Cambridge, MA, USA: MIT
  press, 2016.

\bibitem{pami2013bengio}
Y.~Bengio, A.~Courville, and P.~Vincent, ``Representation learning: {A} review
  and new perspectives,'' \emph{IEEE Trans. Pattern Anal. Mach. Intell.},
  vol.~35, no.~8, pp. 1798--1828, Aug. 2013.

\bibitem{minh2015}
V.~Mnih, K.~Kavukcuoglu, D.~Silver, A.~A. Rusu, J.~Veness, M.~G. Bellemare,
  A.~Graves, M.~Riedmiller, A.~K. Fidjeland, G.~Ostrovski \emph{et~al.},
  ``Human-level control through deep reinforcement learning,'' \emph{Nature},
  vol. 518, no. 7540, p. 529, Feb. 2015.

\bibitem{survey2019}
N.~C. {Luong}, D.~T. {Hoang}, S.~{Gong}, D.~{Niyato}, P.~{Wang}, Y.~{Liang},
  and D.~I. {Kim}, ``Applications of deep reinforcement learning in
  communications and networking: {A} survey,'' \emph{IEEE Commun. Surv.
  Tutor.}, pp. 1--1, to appear 2019.

\bibitem{2014datacenter}
J.~Gao and R.~Jamidar, ``Machine learning applications for data center
  optimization,'' \emph{Google White Paper}, Oct. 27 2014.

\bibitem{DRLSA}
O.~{Naparstek} and K.~{Cohen}, ``Deep multi-user reinforcement learning for
  dynamic spectrum access in multichannel wireless networks,'' in \emph{Global
  Commun. Conf.}, Singapore, Dec 4-8, 2017, pp. 1--7.

\bibitem{DRLMA}
Y.~{Yu}, T.~{Wang}, and S.~C. {Liew}, ``Deep-reinforcement learning multiple
  access for heterogeneous wireless networks,'' in \emph{Intl. Conf. on Comm.},
  Kansas City, USA, May 20 - 24, 2018, pp. 1--7.

\bibitem{DRLHO}
Z.~{Wang}, L.~{Li}, Y.~{Xu}, H.~{Tian}, and S.~{Cui}, ``Handover control in
  wireless systems via asynchronous multiuser deep reinforcement learning,''
  \emph{IEEE Internet Things J.}, vol.~5, no.~6, pp. 4296--4307, Dec 2018.

\bibitem{DRLMS}
Y.~{Sun}, M.~{Peng}, and S.~{Mao}, ``Deep reinforcement learning-based mode
  selection and resource management for green fog radio access networks,''
  \emph{IEEE Internet Things J.}, vol.~6, no.~2, pp. 1960--1971, April 2019.

\bibitem{multiarm2014}
P.~Blasco and D.~G{\" u}nd{\" u}z, ``Learning-based optimization of cache
  content in a small cell base station,'' in \emph{IEEE Intl. Conf. on
  Commun.}, Sydney, Australia, June 10-14, 2014, pp. 1897--1903.

\bibitem{DNN_NonCVX}
L.~{Lei}, L.~{You}, G.~{Dai}, T.~X. {Vu}, D.~{Yuan}, and S.~{Chatzinotas}, ``A
  deep learning approach for optimizing content delivering in cache-enabled
  {HetNet},'' in \emph{Intl. Symp. on Wireless Comm. Systems}, Bologna, Italy,
  August 28--31, 2017, pp. 449--453.

\bibitem{PSN1}
S.~Traverso, M.~Ahmed, M.~Garetto, P.~Giaccone, E.~Leonardi, and S.~Niccolini,
  ``Temporal locality in today's content caching: {W}hy it matters and how to
  model it,'' \emph{ACM SIGCOMM Comput. Commun. Rev.}, vol.~43, no.~5, pp.
  5--12, Nov. 2013.

\bibitem{PSN2}
M.~Leconte, G.~Paschos, L.~Gkatzikis, M.~Draief, S.~Vassilaras, and
  S.~Chouvardas, ``Placing dynamic content in caches with small population,''
  in \emph{Intl. Conf. Comput. Commun.}, San Francisco, USA, April 10-15, 2016,
  pp. 1--9.

\bibitem{RL1}
A.~Sadeghi, F.~Sheikholeslami, and G.~B. Giannakis, ``Optimal and scalable
  caching for {5G} using reinforcement learning of space-time popularities,''
  \emph{IEEE J. Sel. Topics Signal Process.}, vol.~12, no.~1, pp. 180--190,
  Feb. 2018.

\bibitem{RL2}
S.~O. Somuyiwa, A.~Gy{\" o}rgy, and D.~G{\" u}nd{\" u}z, ``A
  reinforcement-learning approach to proactive caching in wireless networks,''
  \emph{IEEE J. Sel. Areas Commun.}, vol.~36, no.~6, pp. 1331--1344, June 2018.

\bibitem{RL3}
A.~Sadeghi, F.~Sheikholeslami, A.~G. Marques, and G.~B. Giannakis,
  ``Reinforcement learning for adaptive caching with dynamic storage pricing,''
  \emph{IEEE J. Sel. Topics Commun.; \emph{see} arXiv:1812.08593}, 2019.

\bibitem{CacheIA}
Y.~{He}, Z.~{Zhang}, F.~R. {Yu}, N.~{Zhao}, H.~{Yin}, V.~C.~M. {Leung}, and
  Y.~{Zhang}, ``Deep-reinforcement-learning-based optimization for
  cache-enabled opportunistic interference alignment wireless networks,''
  \emph{IEEE Trans. Vehicular Tech.}, vol.~66, no.~11, pp. 10\,433--10\,445,
  Nov. 2017.

\bibitem{DRL_AC}
C.~Zhong, M.~C. Gursoy, and S.~Velipasalar, ``A deep reinforcement
  learning-based framework for content caching,'' in \emph{Conf. on Info.
  Sciences and Syst.}, Princeton, NJ, March 21--23, 2018, pp. 1--6.

\bibitem{DRL_Vehic}
Y.~{He}, N.~{Zhao}, and H.~{Yin}, ``Integrated networking, caching, and
  computing for connected vehicles: {A} deep reinforcement learning approach,''
  \emph{IEEE Trans. Veh. Technol.}, vol.~67, no.~1, pp. 44--55, Jan. 2018.

\bibitem{DRL_Smartcity}
Y.~{He}, F.~R. {Yu}, N.~{Zhao}, V.~C.~M. {Leung}, and H.~{Yin},
  ``Software-defined networks with mobile edge computing and caching for smart
  cities: {A} big data deep reinforcement learning approach,'' \emph{IEEE
  Commun. Mag.}, vol.~55, no.~12, pp. 31--37, Dec. 2017.

\bibitem{DRL4edgecaching}
H.~{Zhu}, Y.~{Cao}, W.~{Wang}, T.~{Jiang}, and S.~{Jin}, ``Deep reinforcement
  learning for mobile edge caching: {Review}, new features, and open issues,''
  \emph{IEEE Netw.}, vol.~32, no.~6, pp. 50--57, Nov. 2018.

\bibitem{Maddahali2014}
M.~A. Maddah-Ali and U.~Niesen, ``Fundamental limits of caching,'' \emph{IEEE
  Trans. Inf. Theory}, vol.~60, no.~5, pp. 2856--2867, May 2014.

\bibitem{Distributed2010}
S.~Borst, V.~Gupta, and A.~Walid, ``Distributed caching algorithms for content
  distribution networks,'' in \emph{Intl. Conf. Comput. Commun.}, San Diego,
  CA, USA, Mar. 15-19, 2010, pp. 1--9.

\bibitem{Online2015}
R.~Pedarsani, M.~A. Maddah-Ali, and U.~Niesen, ``Online coded caching,''
  \emph{IEEE/ACM Trans. Netw.}, vol.~24, no.~2, pp. 836--845, Apr. 2016.

\bibitem{collaborative2012}
J.~Dai, Z.~Hu, B.~Li, J.~Liu, and B.~Li, ``Collaborative hierarchical caching
  with dynamic request routing for massive content distribution,'' in
  \emph{Intl. Conf. Comput. Commun.}, Orlando, FL, USA, Mar. 25-30, 2012, pp.
  2444--2452.

\bibitem{decentralized2018}
W.~Wang, D.~Niyato, P.~Wang, and A.~Leshem, ``Decentralized caching for content
  delivery based on blockchain: {A} game theoretic perspective,''
  \emph{arXiv:1801.07604}, 2018.

\bibitem{nygren2010akamai}
E.~Nygren, R.~K. Sitaraman, and J.~Sun, ``The {Akamai} network: {A} platform
  for high-performance {Internet} applications,'' \emph{ACM SIGOPS Operating
  Syst. Rev.}, vol.~44, no.~3, pp. 2--19, 2010.

\bibitem{ramadan2019framework}
E.~Ramadan, P.~Babaie, and Z.-L. Zhang, ``A framework for evaluating caching
  policies in a hierarchical network of caches,'' in \emph{IFIP Networking
  Conference and Workshops}.\hskip 1em plus 0.5em minus 0.4em\relax IEEE, 2019,
  pp. 1--9.

\bibitem{JointDehgan17}
M.~{Dehghan}, B.~{Jiang}, A.~{Seetharam}, T.~{He}, T.~{Salonidis}, J.~{Kurose},
  D.~{Towsley}, and R.~{Sitaraman}, ``On the complexity of optimal request
  routing and content caching in heterogeneous cache networks,'' \emph{IEEE/ACM
  Trans. Netw.}, vol.~25, no.~3, pp. 1635--1648, June 2017.

\bibitem{JointShukla18}
S.~{Shukla}, O.~{Bhardwaj}, A.~A. {Abouzeid}, T.~{Salonidis}, and T.~{He},
  ``Proactive retention-aware caching with multi-path routing for wireless edge
  networks,'' \emph{IEEE J. Sel. Areas Commun.}, vol.~36, no.~6, pp.
  1286--1299, June 2018.

\bibitem{tong2016hierarchical}
L.~Tong, Y.~Li, and W.~Gao, ``A hierarchical edge cloud architecture for mobile
  computing,'' in \emph{IEEE Intl. Conf. on Comput. Commun.}\hskip 1em plus
  0.5em minus 0.4em\relax IEEE, 2016, pp. 1--9.

\bibitem{LTE}
E.~Dahlman, S.~Parkvall, and J.~Skold, \emph{{4G}: {LTE/LTE}-advanced for
  {M}obile {B}roadband}.\hskip 1em plus 0.5em minus 0.4em\relax Academic press,
  2013.

\bibitem{Tsitsiklis}
C.~H. Papadimitriou and J.~N. Tsitsiklis, ``The complexity of {M}arkov decision
  processes,'' \emph{Math. Oper. Res.}, vol.~12, no.~3, pp. 441--450, 1987.

\bibitem{RLbook}
R.~S. Sutton and A.~G. Barto, \emph{Reinforcement {L}earning: {A}n
  {I}ntroduction}.\hskip 1em plus 0.5em minus 0.4em\relax Cambridge, MA: MIT
  press, 2018.

\bibitem{Qlearning}
C.~J. Watkins and P.~Dayan, ``Q-learning,'' \emph{Mach. Learn.}, vol.~8, no.
  3-4, pp. 279--292, May 1992.

\bibitem{fagin1977asymptotic}
R.~Fagin, ``Asymptotic miss ratios over independent references,'' \emph{J.
  Comput. Sys. Sci.}, vol.~14, no.~2, pp. 222--250, 1977.

\bibitem{Conf_expl}
P.~Auer, ``Using confidence bounds for exploitation-exploration trade-offs,''
  \emph{J. Mach. Learn. Res.}, vol.~3, no. Nov, pp. 397--422, 2002.

\bibitem{2016artificial}
S.~J. Russell and P.~Norvig, \emph{Artificial {I}ntelligence: {A} {M}odern
  {A}pproach}.\hskip 1em plus 0.5em minus 0.4em\relax Prentice-Hall, Upper
  Saddle River, NJ, USA, 2016., 2016.

\bibitem{Briefsurveydrl}
K.~Arulkumaran, M.~P. Deisenroth, M.~Brundage, and A.~A. Bharath, ``Deep
  reinforcement learning: {A} brief survey,'' \emph{IEEE Signal Process. Mag.},
  vol.~34, no.~6, pp. 26--38, Nov. 2017.

\bibitem{relus}
G.~{Wang}, G.~B. {Giannakis}, and J.~{Chen}, ``Learning {ReLU} networks on
  linearly separable data: {Algorithm}, optimality, and generalization,''
  \emph{IEEE Trans. Signal Process.}, vol.~67, no.~9, pp. 2357--2370, May 2019.

\bibitem{silver2016}
D.~Silver, A.~Huang, C.~J. Maddison, A.~Guez, L.~Sifre, G.~Van Den~Driessche,
  J.~Schrittwieser, I.~Antonoglou, V.~Panneershelvam, M.~Lanctot \emph{et~al.},
  ``Mastering the game of {Go} with deep neural networks and tree search,''
  \emph{Nature}, vol. 529, no. 7587, p. 484, Jan. 2016.

\bibitem{Machine2016}
Y.~Wu, M.~Schuster, Z.~Chen, Q.~V. Le, M.~Norouzi, W.~Macherey, M.~Krikun,
  Y.~Cao, Q.~Gao, K.~Macherey \emph{et~al.}, ``Google's neural machine
  translation system: {B}ridging the gap between human and machine
  translation,'' \emph{arXiv:1609.08144}, 2016.

\bibitem{lillicrap2016}
T.~P. Lillicrap, J.~J. Hunt, A.~Pritzel, N.~Heess, T.~Erez, Y.~Tassa,
  D.~Silver, and D.~Wierstra, ``Continuous control with deep reinforcement
  learning,'' in \emph{Intl. Conf. on Learn. Representations}, San Juan, Puerto
  Rico, May 2-4, 2016.

\bibitem{dan1990approximate}
A.~Dan and D.~Towsley, \emph{{An Approximate Analysis of the LRU and FIFO
  Buffer Replacement Schemes}}.\hskip 1em plus 0.5em minus 0.4em\relax ACM,
  1990, vol.~18, no.~1.

\bibitem{2018lbc}
B.~Li, T.~Chen, and G.~B. Giannakis, ``Bandit online learning with unknown
  delays,'' in \emph{Intl. Conf. on Artificial Intell. and Stat.}, Naha,
  Okinawa, Japan, April 16-18 2019.

\bibitem{deepcache}
A.~Narayanan, S.~Verma, E.~Ramadan, P.~Babaie, and Z.-L. Zhang, ``Deepcache:
  {A} deep learning based framework for content caching,'' in \emph{Netw. Meets
  AI \& ML}.\hskip 1em plus 0.5em minus 0.4em\relax ACM, Budapest, Hungary,
  August 20-24, 2018, pp. 48--53.

\end{thebibliography}

\end{document}